# Colloidal interactions and unusual crystallization versus de-mixing of elastic multipoles formed by gold mesoflowers

Ye Yuan[1], Mykola Tasinkevych[2,3] & Ivan I. Smalyukh[1,4,5*]

[1]Department of Physics and Soft Materials Research Center, University of Colorado, Boulder, CO 80309, USA

[2]Departamento de Física, Faculdade de Ciências, Universidade de Lisboa, Campo Grande P-1749-016 Lisboa, Portugal

[3]Centro de Física Teórica e Computacional, Universidade de Lisboa, Campo Grande P-1749-016 Lisboa, Portugal

[4]Department of Electrical, Computer, and Energy Engineering, Materials Science and Engineering Program, University of Colorado, Boulder, CO 80309, USA

[5]Renewable and Sustainable Energy Institute, National Renewable Energy Laboratory and University of Colorado, Boulder, CO 80309, USA

*Email: ivan.smalyukh@colorado.edu

***Colloidal interactions in nematic liquid crystals can be described as interactions between elastic multipoles that depend on particle shape, topology, chirality, boundary conditions and induced topological defects. Here, we describe a nematic colloidal system consisting of mesostructures of gold capable of inducing elastic multipoles of different order. Elastic monopoles are formed by relatively large asymmetric mesoflower particles, for which gravity and elastic torque balancing yields monopole-type interactions. High-order multipoles are instead formed by***

*smaller mesoflowers with a myriad of shapes corresponding to multipoles of different orders, consistent with our computer simulations based on free energy minimization. We reveal unexpected many-body interactions in this colloidal system, ranging from de-mixing of elastic monopoles to a zoo of unusual colloidal crystals formed by high-order multipoles like hexadecapoles. Our findings show that gold mesoflowers may serve as a designer toolkit for engineering colloidal interaction and self-assembly, potentially exceeding that in atomic and molecular systems.*

Introduced by Einstein within a theoretical framework explaining Brownian motion of tiny particles[1], colloidal atom paradigm has provided the motivation and means for organizing particles into crystals and other structures, mimicking and even exceeding the diversity of structures in naturally occurring molecular and atomic materials[2]. Long-range elasticity-mediated colloidal interactions between particles[3] in liquid crystal (LC)[4] fluids have enabled a host of anisotropic colloidal self-assemblies and composite materials[5-26]. Colloidal inclusions perturb the uniform background of nematic LC's ground-state unidirectional molecular orientations, producing distortions in the molecular ordering described by the coordinate-dependent director field **n(r)**. These director distortions propagate far beyond the physical extent of the particles themselves[3], though confining surfaces with strong boundary conditions can partially localize and limit the extent of spatial propagation of these distortions[11]. Minimization of director distortions to lower the total free energy cost when colloidal particles are in a close proximity leads to elasticity-mediated interactions not present in isotropic host media[3,11]. Under the one-elastic-constant approximation, the governing Euler-Lagrange equation, derived from minimization of free energy, is of Laplace's type, similar to that of electrostatics, thus allowing for interpreting the nature of long-ranged **n(r)** distortions based on the multipole expansions[3,11,21]. At small inter-particle distances, this multipole description is limited by the presence of topological singularities and non-spherical topographic features of the colloidal particles, whereas surface confinement and boundary conditions on sample surfaces effectively limit this description at large inter-particle distances[11]. The confinement effects effectively screen the long range nature of interactions and can be accounted for analogously to the screening of electrostatic colloidal interactions in presence of counterions and many other types of screening of physical forces described using the mathematical language of multipoles[27-31]. Deviations of **n(r)** in opposite directions away from the far-field uniform alignment can be interpreted analogously to opposite charges in electrostatic charge distributions, defining the design principles for achieving diverse types of colloidal interactions and assemblies that mimic the well-understood interactions between electrostatic charge distributions[11]. In addition to theoretical analysis[5-7,12-15,17], a number of elastic multipoles have been discovered experimentally[3,8-11,16-21]. Surface anchoring boundary conditions on the particles and size, shape, topology and chirality are all found to be important factors, defining behavior of nematic LC colloids[11]. Besides, colloidal particles can induce different multipoles depending on the types of defects that occur. For example, colloidal spheres can induce elastic dipoles[3], quadrupoles[12], or hexadecapoles[17] depending on whether singular point[3] or "Saturn ring" disclination loop[32,33] or simultaneously both types of defects[17] are formed, respectively. However, such ability of achieving different types of elastic multipoles by colloidal inclusions of the same type is limited, consequently limiting the diversity of assemblies and composite materials that can be achieved. Although two-photon polymerization based fabrication of colloidal particles with complex shapes can allow for designing many different types of elastic LC multipoles[11,16], it is limited to particles made of polymers and elastomers and cannot be easily scaled, which is a limitation as compared to wet chemical synthesis of colloids.

Here we describe a nematic colloidal system made from mesoflower colloidal particles[34] capable of inducing elastic multipoles of different order. These mesoflowers are mesostructures of gold with highly diverse shapes and with characteristic sharp spikes of sub-micron dimensions[34]. These complex yet diverse particles allow for inducing different elastic multipoles when dispersed in a nematic LC, which for small particles range from dipoles to hexadecapoles, and even higher order multipoles. Larger asymmetric mesoflowers induce elastic monopoles that emerge due to external gravitational torques/forces that are balanced by those originating from the LC's orientational elasticity; this leads to monopole-type interactions with confining substrates and other colloidal objects, as well as to unusual anisotropy in their Brownian motion. Numerical modeling based on Landau-de Gennes free energy minimization, while accounting for the presence of singular defects with corresponding variations of the scalar order parameter in addition to the director configurations, confirms these experimental observations and predicts the existence of other elastic multipoles achieved by systematically varying the distribution of spikes. We reveal changes in physical behavior like pair interactions of colloidal particles stemming from small changes in particle dimensions and shapes, as exemplified by the mesoflower nematic colloidal system. Numerical modeling of interactions in large colloidal systems shows that these effects lead to unusual types of colloidal crystals and transformations between them in the case of high-order multipoles and to de-mixing of elastic monopoles of opposite signs, very differently from electrostatic interactions in atomic systems. In the spirit of the colloidal atom paradigm, our findings reveal that LC colloids have a great potential of not only expanding the length scales of self-assembly from atomic to colloidal scales[1-3], but also diversifying the forms of colloidal organization by going beyond what is accessible to atomic systems.

## Results

**Experimental generation of elastic multipoles**

Gold mesoflowers of size ranging from hundreds of nanometers to micrometers are synthesized by a seed-mediated growth method (see details in Methods)[34]. They are then dispersed in a nematic LC, 4-cyano-4'-pentylbiphenyl (5CB) (Fig. 1). The cetyltrimethylammonium bromide (CTAB) coating on the surface of the particles sets perpendicular boundary conditions for the director $n(r)$, while the sharp spikes sticking out in all directions perturb the uniform far-field alignment $n_0$ of the LC defined by the rubbing direction of the confining substrates. Although each mesoflower possesses varying number of spikes of different size, the analysis of $n(r)$ around the particles allows for prediction of their colloidal behavior based on electrostatic analogy[11]. In the optical micrographs taken under crossed polarizers with an additional 530 nm phase retardation plate, director distortions manifest themselves as the colored regions different from the background, indicating that $n(r)$ deviates away from $n_0$ around a particle. The direction of director rotations, or rather the rotation of projection of $n(r)$ to the plane of the sample, in blue (yellow) regions of polarizing optical micrographs is extracted on the basis of addition (subtraction) of phase retardation of the waveplate and the birefringent LC sample with the corresponding director orientation patterns. For our experimental geometry, the blue (yellow) polarized interference colors in optical micrographs reveal positive (negative) $x$-components of director and clockwise (counterclockwise) rotations of $n(r)$ away from $n_0$ (see the insets in figures showing the details for particular experiments discussed). The diversity of the mesoflowers leads to a variety of polarized interference color patterns, revealing $n(r)$-distortions resembling that of elastic multipoles[21,26], depending on the exact patterns of inter-changing blue and yellow colored regions distributed around the particles (Fig. 1h-k and Supplementary Fig. 1). In some peculiar cases, the mesoflowers are surrounded by predominantly one color (Figs. 1g and 2a,b), implying $n(r)$ rotation to one direction away from $\mathbf{n}_0$. Such director distortions can be identified as elastic monopoles, although it has long been believed that they should relax to higher order multipoles[11,18,26] because rotations of director in one direction would generate elastic torque that would relax such distortions to minimize free energy. In our system, however, gravity prompts this behavior because the density of gold is much higher than that of the LC and can serve as a source of external torques/forces balancing their elastic counterparts. For sufficiently large particles, gravitational forces and torques can compete with the elastic counterparts and the elastic monopoles can be stabilized rather than relax to higher order multipoles. This behavior is very different from that of conventional colloids in isotropic fluid hosts, where the role of gravity is associated with colloidal particle sedimentation, re-distribution along the sample height or destabilization. As an order of magnitude assessment of these unusual gravity effects in LCs, we equate effective gravitational potential of the mesoflower $\propto\Delta\rho(4\pi R^3/3)gR$ (here $\Delta\rho$ is the difference between densities of gold 19320 kg m$^{-3}$ and the LC 1008 kg m$^{-3}$; $g$=9.8 m s$^{-2}$ is the standard gravity; $R$ is an effective radius of the mesoflower) with the elastic energy $\propto \bar{K}R$ (the used LC's average elastic constant $\bar{K}$~6.5 pN). We obtain an estimate of effective threshold radius for particles significantly influenced by gravitational effects, $R_t$~ 2 µm, well within the range of the studied mesoflower dimensions (Fig. 1). Thus, depending also on particle shape, gravitational effects can be a factor for stabilizing different elastic multipoles when the particle size is comparable or larger than $R_t$. For smaller particles, gravity is not strong enough to compete with elastic forces (Fig. 1d,e), but it can serve as a source of symmetry-breaking torques and forces for particles larger than $R_t$. Particles with intermediate dimensions ~$R_t$ may exhibit monopoles as metastable states due to the interplay between the complex shape of the particle, surrounding director distortions and satellite defects, as well as interactions with the confining substrate. Indeed, when we poke, heat or rotate such particles by laser tweezers, different multipole-like color patterns can appear around the same mesoflower of size $\sim R_t$ (Fig. 2 and Supplementary Fig. 1), including the monopole-like structures.

The intrinsic viscoelastic anisotropy of the LC host leads to anisotropic Brownian motion of colloidal inclusions[35], which further depends on the particle's geometric shape, surface boundary conditions and induced defects[19,36,37]. For mesoflowers with comparable effective dimensions of spikes extending in directions along and perpendicular to $n_0$, the angular dependence of the mean square displacement (MSD) of individual particles is directly related to $n(r)$ structures that they induce (Fig. 2). In an isotropic phase of the LC, like colloidal spheres, mesoflowers diffuse in all directions with roughly equal probability during a long period of time[38], so that the angular dependence of MSD is isotropic. In a nematic phase, even for spherical particles, molecular alignment breaks the symmetry and defines an easy axis for particle diffusion[35], which yields dumbbell-shaped angular dependencies of MSD, aligned along $n_0$ and symmetric with respect to it. However, the mesoflower-induced $n(r)$ distortions further enrich this behavior (Fig. 2). For example, the dumbbell-shaped MSD dependence of elastic monopoles has a long axis tilted away from $n_0$ (Fig. 2a,b), with the tilting direction matching the unidirectionally rotated nearby $n(r)$ orientation and correlating with the elastic monopole sign. In contrast, mesoflowers inducing higher order multipoles exhibit MSD angular dependence symmetric with respect to $n_0$ (Fig. 2c,d and Supplementary Fig. 2). Interestingly, the diffusion behavior can be altered by switching particle-induced structures between different metastable states associated with reconfiguration of $n(r)$ and elastic multipoles (Fig. 2a,c). Such switching of diffusion anisotropy and medium-mediated long-distance correlation of diffusion anisotropy between particles of the same type, cannot be achieved for mesoflowers or other particles dispersed in an isotropic medium, even for anisotropic particles[38].

**Elasticity-mediated colloidal interactions**

Dynamics and interactions of mesoflowers dispersed in the LC are probed under polarizing optical microscopy with a 530 nm retardation plate inserted, so that elastic multipoles can be identified by examining the color patterns. Particles are brought to desired initial positions using laser tweezers and then released. In addition to conventional dipole-dipole (Supplementary Fig. 3) and quadrupole-quadrupole (Fig. 3) interactions that are common in other nematic colloids[11], interactions involving elastic monopoles are also observed (Fig. 4a,b). A monopole-like mesoflower surrounded by unidirectionally rotated **n(r)** (consistent with the blue color in a polarizing micrograph) attracts another mesoflower with a dipolar **n(r)**, surrounded by approximately equal amount of blue and yellow colors within the polarizing micrograph, which is an elastic dipole. The two particles eventually approach each other by sharing blue-colored regions, thus lowering the total energy cost of the ensuing colloidal assembly (Fig. 4a,b). The interaction potential, calculated from the balance of elastic and viscous drag $\propto dr_c/dt$ forces, is in the range of hundreds of $k_BT$ and its power-law distance dependence $\propto -r_c^{-2}$ is consistent with that of the monopole-dipole interaction ($r_c$ is the distance between the centers of interacting colloidal particles). As another example, Fig. 4c,d shows how an assembly of two mesoflowers consisting of one dipole and one monopole attracts another dipolar mesoflower (Fig. 4c,d), with the interactions again consistent with the electrostatic analogy of these nematic colloids. For elastic multipoles of leading orders $2^l$ and $2^m$, the balance of viscous drag and elastic force $\propto 1/r_c^{l+m+2}$ yields the anticipated time dependence of inter-particle distance $r_c(t)=[r_0^{l+m+3}-(l+m+3)\alpha t]^{1/(l+m+3)}$, where $r_0$ is the initial center-to-center distance and $α$ is a fitting parameter, consistent with experiments for pair interactions for all studied multipoles of the same and different orders (Figs. 3, 4 and Supplementary Fig. 3).

**Computer simulations of elastic multipoles and interactions**

Numerical minimization of Landau-de Gennes free energy provides insight into the director configurations around mesoflowers with complex shape and different dimensions (Figs. 5 and 6). These particles induce networks of defect lines meandering on their surfaces, typically along the ridges of spikes, with the long-range distortions of **n(r)** influenced by the spiky topography of particles but not exactly following it (Figs. 5 and 6). For large and strongly asymmetric particles gravitational torques and forces compete with their elastic counterparts, giving the origin to elastic monopoles (Fig. 5), whereas smaller particles tend to induce higher order multipoles (Fig. 6 and Supplementary Fig. 4). For monopole-like particles, the clockwise versus counterclockwise unidirectional rotation of **n(r)** away from **n**$_0$ corresponds to the opposite signs of monopoles, as confirmed by multipole expansion. The *x*-component of director, $n_x$, which is an effective elastic charge density, has the same sign when plotted on a sphere encompassing the monopole-inducing mesoflowers, though its amplitude is nonuniform, as shown in the Fig. 5a-c, consistent with the leading monopole moment of the elastic charge distribution. Since the twist elastic constant of the LC is the smallest, the bend and splay distortions can also relax through equivalent distortions containing twist, so that the particles orientations can deviate away from **n**$_0$ not only in the plane containing **g**, the gravitational acceleration, but also in the plane orthogonal to **g**, as experimentally observed in planar LC cells (Figs. 1-4).

Details of the elastic-gravitational torque balance are revealed by simulating an asymmetric particle (similar to low-symmetry particles that tend to induce elastic monopoles in experiments) placed above a wall in $xz$ plane with tangential boundary conditions (Fig. 5). The $x$-components (see inset in the lower left corner of Fig. 5d for the definition of the reference system) of the elastic $T_x^{el}$ and gravity $T_x^g$ torques vary with the angle $\theta$ between the particle axis (the line connecting the center of the particle core and the tip of one of the spikes) and $\mathbf{n}_0$. The torques are calculated about the center (see Supplementary Fig. 5), and exhibit different signs within the intervals $\theta \in [-180°, -127°]$ and $\theta \in [-90°, 0°]$. The orientations $\theta = 0° (-180°)$ and $\theta \simeq -127°$ correspond to the minima of the elastic and gravity energies, respectively. For particles with masses $M \ll \bar{K}/g$, the elastic energy dominates and the particle adopts one of the two equilibrium orientations either $\theta_{eq} \lesssim 0°$, or $\theta_{eq} \gtrsim -180°$, as shown in Fig. 5f for the case $\theta_{eq} \lesssim 0°$. As the particle size increases, the gravity tilts the particle's axis further away from the purely elastic equilibrium $\theta = 0°$ towards negative values of $\theta_{eq}$, the equilibrium orientation behaves according to the upper branch of the curve in Fig. 5f. When starting from the second elastic energy minimum at $\theta = -180°$, the gravity tilts the axis towards less negative $\theta$ upon increasing the particle size, towards the minimum of the gravitational energy (lower branch of $\theta_{eq}$ curve in Fig. 5f). For sizes larger than $R_t$, the particle would adopt its equilibrium orientation in the vicinity of the purely gravitational equilibrium $\theta \simeq -127°$. The equilibrium orientation $\theta_{eq}$ in Fig. 5f was obtained for $h=3.5R_0$, where $R_0$ is the radius of the spherical core of the mesoflower and $h$ is the distance from the particle center to the wall (Supplementary Fig. 6). As the weight of the particle increases the equilibrium particle-wall separation $h_{eq}$ decreases, as shown in the inset of Fig. 5e (the gravity acts in the directions towards the wall) and depends on balancing of gravitational and the repulsive elastic forces. The latter also depends on the strength of surface anchoring boundary conditions and is calculated here from the free energy profile versus $h$ for the regime of strong boundary conditions corresponding to the experiments. The corresponding free energy results (Fig. 5e) demonstrate particle-wall repulsion, which have similar scaling when using one elastic constant approximation or not, though the elastic torque is almost insensitive to the variation of $h$ for $3.5 \leq h/R_0 \leq 20$, justifying the approximations used for calculating $\theta_{eq}$ (Fig. 5f). The size-dependent levitation of mesoflowers at different $h$ (inset of Fig. 5e) shows how suspensions of mesoflowers in LCs could be potentially used for separating particles of different dimensions, though the range of particle dimensions that can be effectively separated will depend on the density of particles relative to that of the LC host medium. Therefore, this separation method may be limited to high-density particles, like the ones made from gold that we use in this study.

The wall-monopole particle repulsive potential $\propto 1/h$ revealed in Fig. 5e agrees with the predictions of nematostatics[5,6,14] according to which elastic monopoles with opposite signs repel. Indeed, employing the electrostatic analogy, the interaction with the wall can be modeled using an image elastic monopole, yielding the $\propto 1/h$ potential. The potentials of mean force between two monopole particles with the same as well as opposite signs of monopole moments (Fig. 5g) also agree with nematostatics[9]: like (unlike) elastic monopoles attract (repel) each other with an effective potential $\propto 1/r_c$. Numerical modeling provides insights that formation of elastic monopoles by mesoflowers of gold is facilitated not only by their relatively large size $>R_t$, but also by symmetry breaking often caused by asymmetric distribution and dimensions of individual spikes (Supplementary Fig. 5) within the mesoflower particle (both effects enabled by the competition of gravitational forces and torques with their elastic counterparts, which stabilize elastic monopole configurations).

For particles smaller than $R_t$, varying the number and positions of spikes within mesoflowers generates different multipole series of the director distortions (Fig. 6 and Supplementary Fig. 4). The multipole expansion analysis[21] for the numerically simulated **n(r)** shows that certain particles induce stable or metastable structures with strongly pronounced multipole moments of different orders, including octupoles (Fig. 6a-c), hexadecapoles (Fig. 6d,e and Supplementary Fig. 4a-c) and even 64-poles (Fig. 6f,g and Supplementary Fig. 4d-f), with the other multipole moments orders of magnitude smaller. The numerically simulated $n_x$-distributions on spheres encompassing these multipolar mesoflowers are consistent with the corresponding charge distributions of high leading order electrostatic multipoles (Fig. 6), though they also contain fine features dictated by detailed geometry of the mesoflowers (compare Fig. 6a and Fig. 6c).

**Colloidal crystals and de-mixing of elastic multipoles**

A multipolar approximation for an effective interaction between two colloidal particles distance $r_c$ apart and with the center-to-center vector forming an angle $\theta_c$ with the far field director **n**₀ reads[17]:

$$U_{el}(r_c, \theta_c) = 4\pi \bar{K} \Sigma_{l,m} Q_l Q_m (-1)^m (l+m)! \, P_{l+m}(\cos(\theta_c)) \frac{R_{eff}^{l+m+2}}{r_c^{l+m+2}}, \qquad (1)$$

where $P_l(x)$ is the Legendre polynomial of degree $l$, $Q_l$ characterize the strength of elastic multipole moments, and $R_{eff}$ is the characteristic length scale of the multipole (set in our case by the size of the colloidal particle). The analysis of Eq. (1) shows that a large variety of pair interaction patterns arises for multipoles of different orders (Fig. 7), which can lead to the formation of colloidal crystals and other structures arising from the competition between multipolar elastic and screened electrostatic repulsive interactions[19]. Because our mesoflower colloids allow for realization of a large variety of multipoles of different orders, one can systematically explore how such colloidal interaction lead to self-organization depending on the order of the leading-order multipole.

First, we consider a half-half binary mixture of distinct elastic monopoles confined at a plane coplanar to the far field director. Contrary to electric charges, similar elastic monopoles attract and dissimilar repel[9,11], as discussed above for pair interactions. To assess how this behavior impacts collective behavior of many such particles, we assume pairwise additive interaction potential and augment the monopole-monopole elastic interaction with a truncated repulsive Yukawa potential corresponding to the screened electrostatic interactions, which can be tuned by adding counterions through doping LCs with salt and other additives[39]:

$$U_Y(r_c) = \begin{cases} A\, e^{-\kappa r_c} \dfrac{1}{r_c}, & r_c \leq r_{CO} \\ 0, & r_c > r_{CO} \end{cases}, \quad (2)$$

where $r_{CO}$ is the cutoff distance, $A$ and $\kappa$ are positive constants, with the latter characterizing electrostatic screening effects due to counterions within the LC[39]. We emphasize that rather strong electro-static repulsion at short particle separations is needed in order to avoid short-range steric, elastic and van der Waals interactions leading to aggregation due to the complex surface geometries of particles. Assuming such screened electro-static repulsions, we perform molecular dynamics simulations using open source Large-scale Atomic/Molecular Massively Parallel Simulator (LAMMPS)[40]. In this simulation as well as in all the subsequent simulations of self-organizations of higher order elastic multipoles, we neglect possible changes in multipole moments, or generation of additional multipoles, as the inter-particle distances and orientations vary. We also neglect the short-range effects that may arise at small inter-particle distances and that possibly cannot be fully described within the multipole expansion approach that we consider.

Figure 8 shows a series of snapshots along the system trajectory, where a perfect NaCl-like crystal order (see Fig. 8a) is assumed as the initial condition. In the course of time, the system of oppositely charged elastic monopoles undergo a spatial segregation, Fig. 8f, which is the direct consequence of the fact that like (opposite) monopoles attract (repel) (see also the Supplementary Movie 1).

Self-assembly of two-dimensional nematic colloids with the dominant dipolar or quadrupolar elastic interactions has been previously reported[22,41,42] and is consistent with the behavior of colloidal dipoles and quadrupoles formed by gold mesoflowers. Driven by the symmetries of the underlying multipole potentials, the dipole colloids tend to assemble head-to-tail in chains aligned along the far field director. The chains then assemble in a two-dimensional crystalline structure with an anti-ferromagnetic-like alignment of the neighboring chains[22,41]. Quadrupolar colloids reveal two-dimensional crystalline assembly with a rhomboidal unit lattice cell where the colloid-colloid bonds are aligned along the attractive directions of the underlying quadrupole potential[42]. The sectors of attraction and repulsion are oriented with respect to $\mathbf{n}_0$, which defines the long-range orientations for the crystallographic axes of the ensuing colloidal crystals formed by elastic dipoles and quadrupoles, thus precluding formation of grain boundaries between them. To the best of our knowledge, crystallization of nematic colloids driven by higher order multipole potentials have not been investigated.

Experimental results[20-22,41,42] suggest that the symmetry of the eventual nematic colloids crystals are mainly determined by the distribution of the attractive directions of the corresponding multipole potential. The elastic octupole potential has six directions of attraction forming roughly 60° between each other (Fig. 7a). It is therefore expected that octupole nematic colloids crystalize in two dimensions into a hexagonal lattice, indeed consistent with our LAMMPS-based simulations. The unique alignment of the hexagonal lattice of these elastic-octupole-based crystals with respect to $\mathbf{n}_0$ in this nematic colloidal system precludes formation of grain boundaries and, because of this, could potentially be used for the formation of large crystal lattices. The case of hexadecapole colloids is fundamentally more interesting, as in this case the potential has eight directions of attractions (Fig. 7b), which permits two types of quadratic lattice arrangements of hexadecapole colloidal particles, with the corresponding unit lattice cells shown in Fig. 9a,b. In order to verify this hypothesis, we performed molecular dynamic simulations (using LAMMPS) of two-dimensional hexadecapole colloids, regularized with repulsive Yukawa potential as defined by Eq. (2). We used constant temperature and constant pressure (NPT) ensemble. Initially the system was subject to a strong isotropic pressure (Fig. 9c and f). In this initial state the colloids formed a glassy structure, with enhanced local hexagonal ordering revealed by the absolute value $|q_6|$ of the local hexatic bond order parameter (Fig. 9c) and suppressed $|q_4|$ of the quartic bond order parameter (Fig. 9f). In the course of the simulations, the external isotropic pressure was gradually decreased, which resulted in spontaneous formation of two types of quadratic colloidal lattices and associated grain boundaries between them (Fig. 9d, e, g and h, and Supplementary Movie 2).

Although the hexadecapole elastic interactions tend to form two-fold degenerate ground state whose lattice units are depicted in Fig. 9a,b, this tendency can be inconsistent with the efficiency of two-dimensional packing at high colloidal concentrations. Colloidal particles are electrostatically charged and mutually repel with screened Coulomb electrostatic potential, where the range and the strength of this potential can be tuned[39], which provides additional knobs to control the self-assembly of nematic colloids. We have performed illustrative molecular dynamic simulations of self-assembly behavior of hexadecapole colloids with strong, weakly screened repulsive Yukawa potential. The ensuing self-assembled colloidal configurations are presented on Fig. 10 and Supplementary Fig. 7 (see also Supplementary Movies 3-6). At high concentrations the system reveals glassy (Fig. 10a,d) configurations, while at low concentrations particles assemble into repulsion-induced sparse hexagonal lattices (not shown). Surprisingly, at intermediate values of particle concentrations and depending on the strength of the hexadecapole elastic moment, the system can crystalize into a range of rhomboidal lattices, with some examples shown in Fig. 10c, k and Supplementary Fig. 7c, k, and the corresponding unit lattices in Fig.10q. Importantly, the symmetry of these lattices differs from the ones of hexadecapolar ground states shown in Fig. 9a,b. Additionally, at some number densities we observe coexisting rhomboidal and hexagonal lattices, see Fig. 10c, k and Supplementary Fig. 7c, k,. Many interesting energetically comparable colloidal organizations emerge as a result of dense packing and elastic hexadecapolar interactions favoring different lattices. This unusual crystallization behavior is revealed by analyzing bond orientational order parameters within self-assembled crystallites separated by grain boundaries. These examples illustrate the ability of hexadecapolar nematic colloids to self-assemble into a range of two-dimensional low-symmetry crystal structures, with properties that can be designed through controlling the particle shape, concentration as well as internal properties of the nematic host.

**Discussion**

In this work, we have developed a nematic colloidal system of mesoflowers where complex and diverse shapes of mesostructures induce elastic multipoles of different order, from monopoles to high-order multipoles like 64-poles. By varying the basic building parts of these particles, their colloidal behavior can be effectively controlled. As the shape and size of particles vary, they behave like elastic monopoles that (differently from their electrostatic counterparts) tend to de-mix into the domains of like-charged elastic monopoles, or as high-order elastic multipoles (e.g. hexadecapoles) forming unusual crystals that can be tuned by packing density and strength of electrostatic repulsions. While here we focused only on the analysis of two-dimensional crystal lattices for illustrative purposes, even more complex behavior is expected in three dimensions, where even elastic quadrupoles can form triclinic crystal lattices[39]. As the symmetries of elastic potentials of high-order multipoles, which form the basis of crystals, become intrinsically incompatible with crystalline lattices, various forms of frustration can arise and lead to complex crystallization behavior that can be potentially tuned through changing concentration of counterions and surface charging. In addition to crystals, quasi-crystals and various plastic crystals (with positional order but lacking orientational ordering) could possibly also arise due to diverse multipolar nature of our mesoflower colloids, whereas polydisperse systems are expected to form disordered glassy states. Our findings suggest that these mesostructures may serve as a designer toolkit for engineering pre-defined colloidal interaction and self-assembly. Access to these structures in large quantities through chemical synthesis may facilitate developments of new composite materials fabricated using colloidal self-assembly. On the other hand, the sharp geometric features of the gold mesoflowers within these colloidal particle assemblies may be of interest for plasmonic enhancement and related applications.

## Methods

**Sample preparation**

Gold mesoflowers are synthesized following an aqueous seed-mediated growth procedure[34]. First, 25 mg of citric acid and 1 mL of 25 mM $HAuCl_4$ solution were combined in 35 mL deionized water at 80 °C. Immediately after the color of the solution changed from pale yellow to pink, 100 µL of aniline and 0.5 mL of 25 mM $HAuCl_4$ solution were injected with temperature maintained for 5 more minutes. After 5 hrs, the resultant solution was centrifugated at 4000 rpm to obtain the supernatant as the seed solution for the following steps. The growth solution was prepared by mixing 20 mL of 0.1 M CTAB solution, 335 µL of 25 mM $HAuCl_4$ solution, 125 µL of 10 mM $AgNO_3$ solution and 135 µL of 0.1 M ascorbic acid solution at 80 °C. 2 mL of the seed solution was added to this growth solution, accompanied with gentle mixing. The mixture was kept at 80 °C for 1 hr and cooled down naturally before taking it out for centrifugation at 4000 rpm for 5 mins. The residue was collected and washed in deionized water 3 times and then dried in an oven. Liquid crystal suspension of mesoflowers was obtained by sonicating these dried residues within the LC host, after which the suspension was infiltrated between glass substrates with gap defined by monodisperse glass spacers of 10-20 µm in diameter. Prior to being used in cell preparation, the glass substrates were spin-coated with polyimide (PI2555, from HD Microsystem) and rubbed unidirectionally to define the bulk LC alignment. The cell edges were sealed using fast-setting epoxy glue.

**Optical video microscopy and laser trapping**

An inverted optical microscope (IX81, Olympus) with a charge-coupled device (CCD) camera (Flea, PointGrey) and a holographic laser trapping system operating at 1064 nm was used to take polarizing optical micrographs and videos of mesoflowers in the LC, to probe the stable and metastable director configurations around the particles, as well as to set the initial conditions for studying their pair interactions. The laser was turned off during particle diffusion and interaction when videos were being recorded. To assure good optical imaging resolution and robust laser manipulations, we used an oil-immersion objective lens (UPlanSApo 100×, Olympus) with high numerical aperture NA=1.40 for both imaging and laser trapping. In order to analyze colloidal interactions and diffusion in directions along and perpendicular to $\mathbf{n}_0$ in the plane of the LC cell, optical videos were analyzed using an image-processing software (ImageJ, freeware from the National Institute of Health) to extract positions of the mesoflowers on a frame-by-frame basis. MSDs that characterize single particle's Brownian motion were calculated by analyzing particle displacements along directions perpendicular and parallel to $\mathbf{n}_0$ (denoted $\Delta r_x$ and $\Delta r_z$, respectively) based on the particle's positions in each frame of the video. To probe the angular dependence of particle diffusion, a rotation operation was applied

$$\begin{pmatrix} \Delta r'_x \\ \Delta r'_z \end{pmatrix} = \begin{pmatrix} \cos\beta & \sin\beta \\ -\sin\beta & \cos\beta \end{pmatrix} \begin{pmatrix} \Delta r_x \\ \Delta r_z \end{pmatrix}, \quad (3)$$

where $\beta$ is the angle of rotation with respect to $\mathbf{n}_0$ and $\begin{pmatrix} \Delta r'_x \\ \Delta r'_z \end{pmatrix}$ is the displacements in the rotated coordinate frames. MSD was then calculated as $\langle \Delta r_x \rangle^2$ for $\beta$ varying from 0° to 360°. Diffusion constants $D$ were extracted from MSDs using the relation $\langle \Delta r_x \rangle^2 = 2D\tau$ where $\tau=1/f$ and $f=15$ Hz is the frame rate of the experimental videos. From Stokes-Einstein relation, the drag coefficients of diffusing particles were calculated as $c_\mu = k_B T/D_\mu$ ($\mu=x, z$) and the corresponding drag forces as $F_d = cv$, where $k_B$ is the Boltzmann constant, $T=300$ K is the room temperature and $v$ is the linear speed of the particle. Because of the highly overdamped nature of our colloidal system, the elastic interaction forces could be then found from their balance with $F_d$. The elastic interaction potential was then calculated by integrating the elastic force over distance.

**Simulation of director structures and elastic interactions**

The generation of elastic monopole and higher-order multipoles in the **n(r)** field by mesoflowers is also approached computationally, via numerical minimization of the Landau-de Gennes free energy[26]. The Landau-de Gennes free energy functional depends on the tensor order parameter field **Q(r)** and its spatial derivatives. The functional combines terms which account for variable degree of nematic order, nematic elasticity and biaxiality (at the cores of topological defects), as well as surface anchoring. As such, the Landau-de Gennes approach yields theoretical characterization of **n(r)** and local changes in the degree of nematic ordering that correspond to global or local minima of the free energy functional[26]. Spiky particles are constructed as a union of a spherical core and a given number of spikes distributed over the surface of the core at predefined locations and orientations (Supplementary Fig. 6). Minimization of the free energy is performed numerically for finite homeotropic surface anchoring at the particle surfaces, and by using variable three-dimensional tetrahedral grids[43]. This minimization yields stable or metastable **n(r)** structures around the particles, which are then compared to the experimentally reconstructed counterparts[16,21].

Nematic configurations around mesoflowers are obtained via numerical minimization of the phenomenological Landau-de Gennes free energy functional[26]

$$F_{\text{LdG}} = \int_V \left( a Q_{ij}^2 - b Q_{ij} Q_{jk} Q_{ki} + c \left( Q_{ij}^2 \right)^2 + \frac{L_1}{2} \partial_k Q_{ij} \partial_k Q_{ij} + \frac{L_2}{2} \partial_j Q_{ij} \partial_k Q_{ik} \right) dV + W \int_{\partial V} f_s(Q_{ij}) \, dS \quad (4)$$

where $Q_{ij} = Q_{ji}$ ($i, j = 1, \ldots, 3$) is a traceless tensor order parameter and summation over repeated indices is assumed. In equation (4), the parameter $a$ (unlike the constants $b$ and $c$) is assumed to depend linearly on temperature $T$: $a(T) = a_0(T - T^*)$, where $a_0$ is a material dependent constant, and $T^*$ is the supercooling temperature of the isotropic phase. Phenomenological parameters $L_1$ and $L_2$ are related (via an uniaxial Ansatz for $Q_{ij}$) to the Frank-Oseen elastic constants. We describe finite homeotropic anchoring, with the strength coefficient $W$, at the surface of the particles by using $f_s(Q_{ij}) = (Q_{ij} - Q_{ij}^s)^2$, with $Q_{ij}^s = \frac{3Q_b}{2}(\nu_k \nu_j - \frac{\delta_{kj}}{3})$, where $\delta_{ij}$ is the Kronecker delta symbol, and $\boldsymbol{\nu}$ is the unit outward normal vector to the particle surface[44]; $Q_b = b/8c\,(a + \sqrt{1 - 64\,ac/(3b^2)})$ is the value of the scalar order parameter in the nematic phase, which is thermodynamically favored for $24ac/b^2 < 1$. Minimization of the free energy Eq. (4) is then performed numerically by employing adaptive mesh finite elements method as described in more details in Ref. 43. This minimization yields theoretical characterization of $\mathbf{n}(\mathbf{r})$ and local changes in the degree of order that correspond to global or local minima of the free energy[26].

In our calculations, we use $a_0 = 0.044 \times 10^6$ J m$^{-3}$, $b = 0.816 \times 10^6$ J m$^{-3}$, $c = 0.45 \times 10^6$ J m$^{-3}$, $L_1 = 6 \times 10^{12}$ J m$^{-1}$, and $L_2 = 12 \times 10^{12}$ J m$^{-1}$, which are typical values for 5CB[45] at $T^* = 307$ K. For these values of the model parameters, the bulk correlation length $\xi = 2\sqrt{2c(3L_1 + 2L_2)}/b \approx 15$ nm at the isotropic-nematic coexistence[46], $24ac/b^2 = 1$. We use $W = 5 \times 10^{-4}$ J m$^{-2}$, or $10^{-4}$ J m$^{-2}$, corresponding to the surface anchoring strengths of interactions for substrates coated with CTAB[47-49].

**Computational geometry of colloidal mesoflowers**

We exploit the Open Source Gnu Triangulated Surface (GTS) library[50] to create triangulated surfaces of the mesoflowers. A spiky mesoflower particle is constructed as a union of a spherical core with the radius $R_0$ and a given number of spike particles (Supplementary Fig. 6) decorating the spherical core at predifined locations and orientations. Each spike is chosen as a generalized cylinder with a 5-fold symmetry axis, consistent with the experimental electron microscopy images.

The lateral surface of each spike whose symmetry axis coincides with the z-axis of a Carthesian coordinates system $(x_i, y_i, z_i), i = 1,\ldots,5$, is parametrized as follows:

$$\begin{cases} x_i(s,u) = A(u)[1 + 0.3\cos(5s)]\cos(s), \\ y_i(s,u) = A(u)[1 + 0.3\cos(5s)]\sin(s), \quad (5) \\ \phantom{xxx} z_i(s,u) = Hu, \end{cases}$$

where $0 \leq s < 2\pi$, and $0 \leq u \leq 1$ are parameters; $A(u) = A_0 + (A_0 - A_1)u$ accounts for the variation of the spike "radius" along the spike symmetry axis with $A_0 > A_1$, and $H$ is the height of the spike along its symmetry axis.

The triangulated surface of the spike can be manipilated using the functions implemented in the GTS library, including surface translation, rotation, or rescaling. By using these functions together with the GTS function which merges any two surfaces, we generated the spiky mesoflower particles with arbitrary number of spikes placed at predefind relative positions and with predefined orientations. The triangulation of the nematic domain was then performed by using a quality tetrahedral mesh generator[51], which supports adaptive mesh refinement. Finaly, the discretized functional (5) is minimized numerically using INRIA's M1QN3 optimization routine[52], which implements a limited-memory quasi-Newton method[53].

**Molecular dynamics simulations**

The self-assembly and formation of colloidal crystals are computer-simulated using an open source Large-scale Atomic/Molecular Massively Parallel Simulator (LAMMPS)[40]. In the simulations, we use $\overline{K}R_{\text{eff}}$ as the unit energy, and $R_{\text{eff}}$ as the unit length. The total number of particles simulated is $N\sim 1,000$, with the system size $L_x \times L_y$ ranging from 26×26 to 37×37 $R^2_{\text{eff}}$. The potential energy of the system is assumed to be pair-wise addative, combining both the elastic energy and Yukawa potential: $U=U_{\text{el}}+U_{\text{Y}}$. For the computer simulations on the segregation of monopoles (Fig. 8), the system size is 34×34 $R^2_{\text{eff}}$ and total number of particles $N=1024$ with half of them having positive elastic charges and the other half having negative elastic charges. The interaction potential between particles is taken as

$$U_1(r_c) = 4\pi \bar{K} Q_1^a Q_1^b \frac{R_{\text{eff}}^2}{r_c} + +U_Y(r_c) \quad (6)$$

where the elastic monopole charges $Q_1^a, Q_1^b = \pm 0.1$, and we set $\frac{r_{co}}{R_{\text{eff}}} = 1.5$, $\kappa R_{\text{eff}} = 2$, $\frac{A}{\bar{K} R_{\text{eff}}^2} = 0.1$. The simulation is performed using *NVT* ensemble. Temperature $T$ is fixed by applying a Langevin thermostat with $\frac{k_B T}{\bar{K} R_{\text{eff}}} = 10^{-3}$. For the computer simulations involving hexadecapoles (Figs 9, 10, and Supplementary Fig. 7), $N=855$ and the effective pair interaction potential is

$$U_4(r_c, \theta_c) = 4 \times 8! \pi \bar{K} Q_4^2 P_8(\cos(\theta_c)) \frac{R_{\text{eff}}^{10}}{r_c^9} + U_Y(r_c) \quad (7)$$

The elastic hexadecapolar moment $Q_4$, and the relative Yukawa potential strength $\frac{A}{\bar{K} R_{\text{eff}}^2}$ are varied to obtain different colloidal assembly structures. The simulations are performed in *NPT* ensemble, using a Nose/Hoover temperature thermostat with $\frac{k_B T}{\bar{K} R_{\text{eff}}} = 10^{-3}$, and Nose/Hoover pressure barostat. In order to achieve a spontaneous crystallization an isotropic external pressure is gradually released in the course of the simulations leading to the expansion of the system. The local quartic (hexatic) orientational order parameter $q_4(j)$ ($q_6(j)$), where $j$ is the index of the particle under consideration, is defined as $q_n(j) = \frac{1}{n}\Sigma_{k=1}^n \exp(in\theta_{jk})$; $\theta_{jk}$ is the angle between the particles center-to-center vector $\mathbf{r}_{jk}$ and $x$ axis, and the sum runs over $n$ nearest neighbors of particle $j$.

**References**

1. Anderson, V. J. & Lekkerkerker, H. N. W. Insights into phase transition kinetics from colloidal science. *Nature* **416**, 811-815 (2002).

2. Manoharan, V. N. Colloidal matter: packing, geometry, and entropy. *Science* **349**, 1253751 (2015).

3. Poulin, P., Stark, H., Lubensky, T. C. & Weitz, D. A. Novel colloidal interactions in anisotropic fluids. *Science* **275**, 1770-1773 (1997).

4. Chaikin, P. M. & Lubensky, T. C. *Principles of Condensed Matter Physics* (Cambridge University Press, Cambridge, 2000).

5. Pergamenshchik, V. M. & Uzunova, V. A. Coulomb-like interaction in nematic emulsions induced by external torques exerted on the colloids. *Phys. Rev. E* **76**, 011707 (2007).

6. Pergamenshchik, V. M. & Uzunova, V. A. Elastic charge density representation of the interaction via the nematic director field. *Eur. Phys. J. E* **23**, 161-174 (2007).

7. Lev, B. I., Chernyshuk, S. B., Tomchuk, P. M. & Yokoyama, H. Symmetry breaking and interaction of colloidal particles in nematic liquid crystals. *Phys. Rev. E* **65**, 021709 (2002).

8. Lee, B., Kim, S., Kim, J. & Lev, B. Coulomb-like elastic interaction induced by symmetry breaking in nematic liquid crystal colloids. *Sci. Rep.* **7**, 15916 (2017).

9. Yuan, Y., Liu, Q., Senyuk, B. & Smalyukh, I. I. Elastic colloidal monopoles and reconfigurable self-assembly in liquid crystals. *Nature* **570**, 214-218 (2019).

10. Lapointe, C. P., Mason, T. G. & Smalyukh, I. I. Shape-controlled colloidal interactions in nematic liquid crystals. *Science* **326**, 1083-1086 (2009).

11. Smalyukh, I. I. Liquid Crystal Colloids. *Annu. Rev. Condens. Matter Phys.* **9**, 207–226 (2018).

12. Stark, H. Physics of colloidal dispersions in nematic liquid crystals. *Phys. Rep.* **351**, 387-474 (2001).

13. Lubensky, T. C., Pettey, D., Currier, N. & Stark, H. Topological defects and interactions in nematic emulsions. *Phys. Rev. E* **57**, 610–625 (1998).

14. Pergamenshchik, V. M. & Uzunova, V. A. Colloidal nematiostatics. *Condens. Matter. Phys.* **13**, 1-29 (2010).

15. Tovkach, O. M., Chernyshuk, S. B. & Lev, B. I. Theory of elastic interaction between arbitrary colloidal particles in confined nematic liquid crystals. *Phys. Rev. E* **86**, 061703 (2012).

16. Yuan, Y., Martinez, A., Senyuk, B., Tasinkevych, M. & Smalyukh, I. I. Chiral liquid crystal colloids. *Nat. Mater.* **17**, 71-78 (2018).

17. Senyuk, B., Puls, O., Tovkach, O., Chernyshuk, S. & Smalyukh, I. I. Hexadecapolar nematic colloids. *Nat. Commun.* **7**, 10659 (2016).

18. Brochard, F. & de Gennes, D. G. Theory of magnetic suspensions in liquid crystals. *J. Phys.* **31**, 691-708 (1970).

19. Mundoor, H., Senyuk, B. & Smalyukh, I. I. Triclinic nematic colloidal crystals from competing elastic and electrostatic interactions. *Science* **352**, 69-73 (2016).

20. Mundoor, H., Park, S., Senyuk, B., Wensink, H. H. & Smalyukh, I. I. Hybrid molecular-colloidal liquid crystals. *Science* **360**, 786-771 (2018).

21. Senyuk, B., Aplinc, J., Ravnik, M., & Smalyukh, I. I. High-order elastic multipoles as colloidal atoms. *Nat. Commun.* **10**, 1825 (2019).

22. Muševič, I., Škarabot, M., Tkalec, U., Ravnik, M. & Žumer, S. Two-dimensional nematic colloidal crystals self-assembled by topological defects. *Science* **313**, 954-958 (2006).

23. Phillips, P. M., Mei, N., Reven, L. & Rey, A. Faceted particles embedded in a nematic liquid crystal matrix: Textures, stability and filament formation. *Soft Matter* **7**, 8592-8604 (2011).

24. Hashemi, S. M. et al. Fractal nematic colloids. *Nat. Commun.* **8**, 14026 (2017).

25. Solodkov, N., Shim, J. & Jones, J. C. Self-assembly of fractal liquid crystal colloids. *Nat. Commun.* **10**, 198 (2019).

26. de Gennes, P. G. *The Physics of Liquid Crystals* (Oxford University Press, New York, 1974).

27. Jackson, J. D. *Classical Electrodynamics* (John Wiley & Sons, Inc., New York, 1962).

28. Petschulat, J. *et al*. Multipole approach to metamaterials. *Phys. Rev. A* **78**, 043811 (2008).

29. Thorne, K S. Multipole expansions of gravitational radiation. *Rev. Mod. Phys*. **52,** 299–339 (1980).

30. Born, M. *Atomic Physics* 8th ed (Dover Publications, Mineola, 1989).

31. Cosgrove, T. *Colloidal Science: Principles, Methods and Applications* (Blackwell Publishing, Oxford, 2005).

32. Ruhwandl R. W. & Terentjev E. M. Long-range forces and aggregation of colloid particles in a nematic liquid crystal. *Phys. Rev. E* **55**, 2958-2961 (1997).

33. Gu, Y., & Abbott, N. L. Observation of Saturn-ring defects around solid microspheres in nematic liquid crystals. *Phys. Rev. Lett.* **85**, 4719-4722 (2000).

34. Sajanlal, P. R. & Pradeep, T. Mesoflowers: a new class of highly efficient surface-enhanced Raman active and infrared-absorbing materials. *Nano Res.* **2**, 306-320 (2009).

35. Loudet, J. C., Hanusse, P. & Poulin, P. Stokes drag on a sphere in a nematic liquid crystal. *Science* **306**, 1525 (2004).

36. Senyuk, B., Glugla, D. & Smalyukh, I. I. Rotational and translational diffusion of anisotropic gold nanoparticles in liquid crystals controlled by varying surface anchoring. *Phys. Rev. E*

**88**, 062507 (2013).

37. Koenig Jr., G. M. et al. Single Nanoparticle Tracking Reveals Influence of Chemical Functionality of Nanoparticles on Local Ordering of Liquid Crystals and Nanoparticle Diffusion Coefficients. *Nano Lett.* **9**, 2794-2801 (2009).

38. Han, Y., Alsayed, A. M., Nobili, M., Zhang, J., Lubensky, T. C. & Yodh, A. G. Brownian motion of an ellipsoid. *Science* **314**, 626-630 (2006).

39. Mundoor, H. et al. Electrostatically controlled surface boundary conditions in nematic liquid crystals and colloids. *Sci. Adv.* **5**, eaax4257 (2019).

40. Plimpton, S. Fast parallel algorithms for short-range molecular dynamics. *J. Comp. Phys*. **117**, 1-19 (1995).

41. Škarabot, M. *et al*. Two-dimensional dipolar nematic colloidal crystals. *Phys. Rev. E* **76**, 051406 (2007).

42. Škarabot, M. *et al*. Interactions of quadrupolar nematic colloids. *Phys. Rev. E* **77**, 031705 (2008).

43. Tasinkevych, M., Silvestre, N. M. & Telo da Gama, M. M. Liquid crystal boojum-colloids. *New J. Phys.* **14**, 073030 (2012).

44. Nobili, M. & Durand, G. Disorientation-induced disordering at a nematic-liquid-crystal–solid interface. *Phys. Rev. A* **46**, R6174-R6177 (1992).

45. Kralj, S., Žumer, S. & Allender, D. W. Nematic-isotropic phase transition in a liquid crystal droplet. *Phys. Rev. A* **43**, 2943–2952 (1991).

46. Chandrasekhar, S. *Liquid Crystals*, 2nd ed. (Cambridge University Press: Cambridge, 1992).

47. Proust, J. E., Ter-Minassian-Saraga, L. & Guyon, E. Orientation of a nematic liquid crystal by suitable boundary surfaces. *Solid State Commun.* **11**, 1227-1230 (1972).

48. Stamatoiu, O., Mirzaei, J., Feng, X. & Hegmann, T. Nanoparticles in Liquid Crystals and Liquid Crystalline Nanoparticles. *Top. Curr. Chem.* **318**, 331-394 (2012).

49. Bezrodna, T. *et al*. Structure peculiarities and optical properties of nanocomposite: 5CB liquid crystal–CTAB-modified montmorillonite clay. *Liq. Cryst.* **37**, 263-270 (2010).

50. *GNU Triangulated Surface Library*. Available at http://gts.sourceforge.net (2006).

51. Si, H. *TetGen: a quality tetrahedral mesh generator and a 3D delaunay triangulator*. Available at http://wias-berlin.de/software/tetgen/ (2011).

52. Gilbert, J. C. & Lemaréchal, C. Some numerical experiments with variable-storage quasi-Newton algorithms. *Math. Program*. **45,** 407-435 (1989).

53. Nocedal, J. Updating quasi-Newton matrices with limited storage. *Math. Comput*. **35**, 773-782 (1980).

**Code availability.** All codes used in this work are freely available from the authors upon a request.

**Data availability.** The data that support the findings of this study are available from the corresponding author on request.

**Acknowledgements.** We thank B. Senyuk, B. Lev, S. Chernyshuk and V. Pergamenshchik for discussions and A. Sanders for technical assistance. This research was supported by the U.S. Department of Energy, Office of Basic Energy Sciences, Division of Materials Sciences and Engineering, under contract DE-SC0019293 with the University of Colorado at Boulder (Y.Y. and I.I.S.). Publication of this article was partially funded by the University of Colorado Boulder Libraries Open Access Fund. M.T. acknowledges financial support from the Portuguese Foundation for Science and Technology (FCT) under Contracts No. IF/00322/2015, UID/FIS/00618/2019 and PTDC/FIS-MAC/28146/2017 (LISBOA-01-0145-FEDER028146).

**Author Contributions.** Y.Y. and I.I.S. conducted experimental work and analyzed data. M.T. performed numerical modeling. Y.Y., M.T. and I.I.S. wrote the manuscript. I.I.S. conceived, designed and directed the project.

**Additional information.** Supplementary information is available in the online version of the paper. Reprints and permissions information is available at www.nature.com/reprints. Correspondence and requests for materials should be addressed to I.I.S.

**Competing interests.** The authors declare no Competing Financial or Non-Financial Interests.

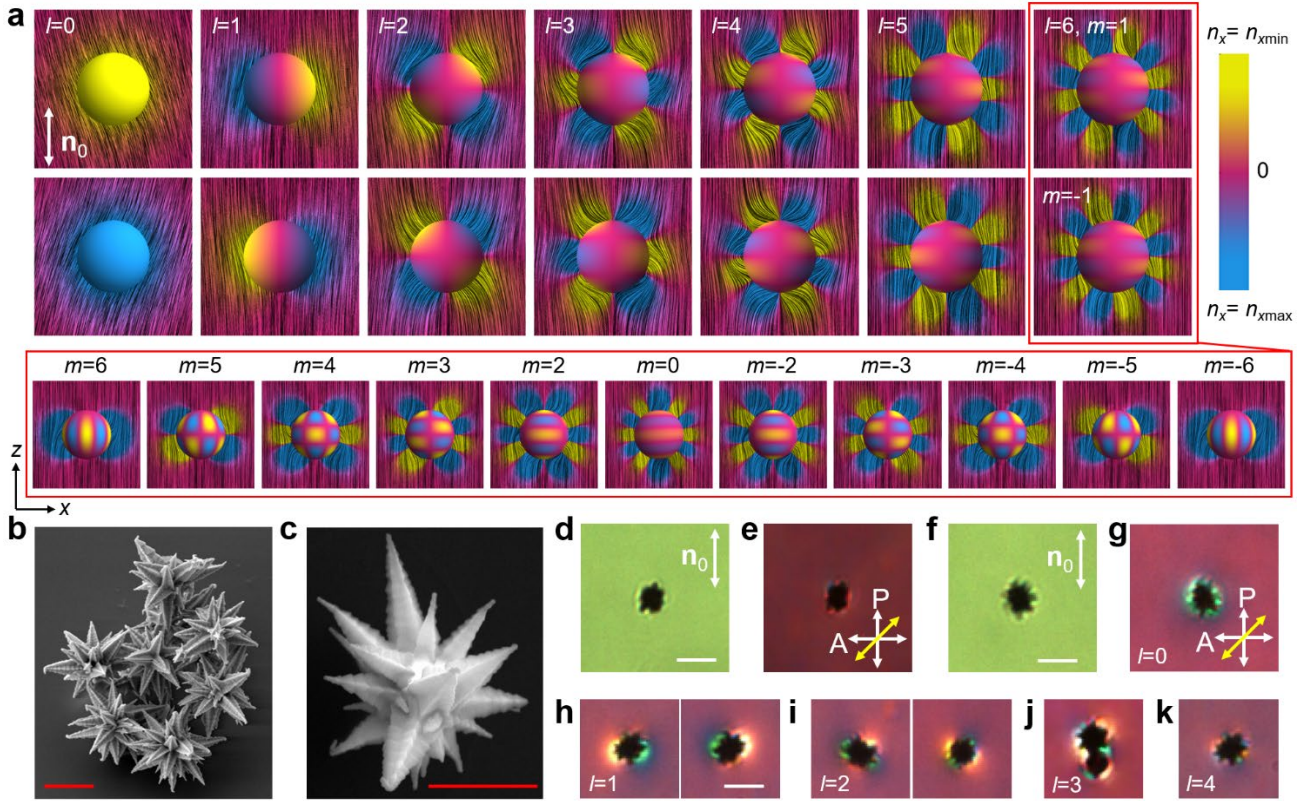

**Fig. 1 | Elastic multipoles generated by gold mesoflowers in a nematic LC. a**, Pure elastic multipoles that can be induced by spherical colloidal surfaces with pre-defined boundary conditions. Each schematic shows the director field **n(r)** in $xz$ plane corresponding to multipole expansions of $l$=1-6 with $m$=1 (top row) and $m$=-1 (bottom row), and for $l$=0 the sign of $n_x$ defines the color; $z$ axis is defined to be along $\mathbf{n}_0$. Blue, yellow and magenta colors indicate positive, negative, and near-zero $n_x$, as shown by the color scale bar on the right. The bottom inset represents pure elastic multipoles with $l$=6 and all values of $m$ that these surfaces can induce. We emphasize that the opposite signs of monopoles have opposite signs of $n_x$ and all other multipoles of opposite signs have opposite alternations of distorted regions around particles with positive and negative $n_x$. **b,c**, Scanning electron microscope images of gold mesoflowers featuring many sharp spikes (**b**) with star-shaped cross-sections (**c**). Red scale bars are 1μm. **d,e**, Bright-field (**d**) and polarizing (**e**) optical micrographs of a smaller mesoflower dispersed in a uniformly aligned LC. $\mathbf{n}_0$ is indicated by the white double arrow; P and A show the crossed polarizer and analyzer; yellow double arrow shows slow axis of a 530 nm retardation plate inserted between the polarizers. **f,g**, Bright-field (**f**) and polarizing (**g**) optical micrographs of a larger mesoflower embedded in a uniformly aligned LC. **h**, Polarizing optical micrographs of mesoflowers generating dipolar ($l$=1) director distortions, with the opposite dipole directions revealed by the color patterns in micrographs on the right and left sides. **i**, Polarizing optical micrographs of mesoflowers generating quadrupoles ($l$=2) of opposite sign. **j,k**, Generation of higher-order elastic multipoles such as octupole ($l$=3) (**j**) due to an assembly of a mesoflower and colloidal sphere and hexadecapole ($l$=4) (**k**) induced by an individual particle. Micrographs **h-k** are taken under the same condition as **e** and **g**. White scale bars are 3 μm.

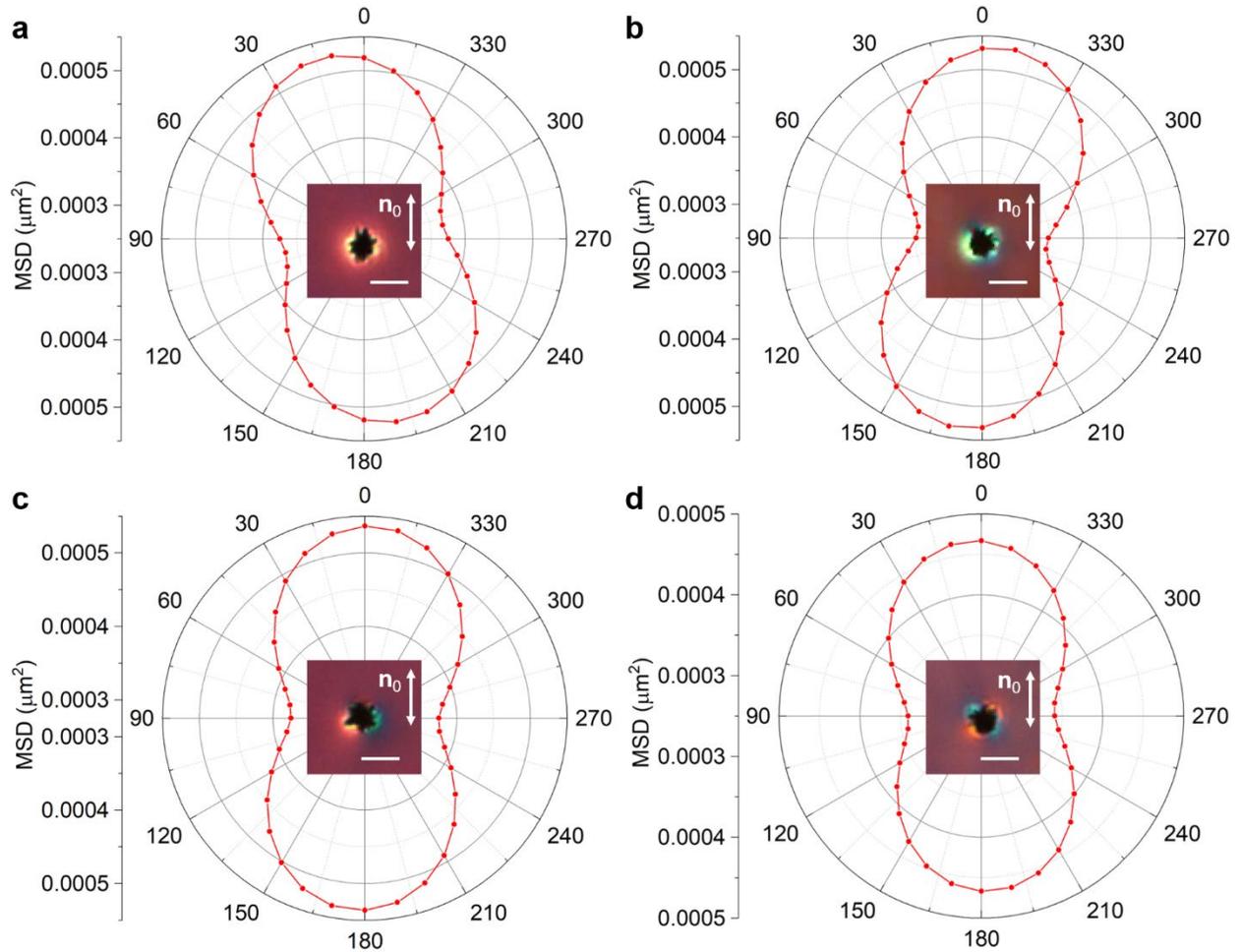

**Fig. 2 | Anisotropic Brownian motion of mesoflowers inducing different elastic multipoles.** **a,b**, Angular-dependent mean square displacement (MSD) of mesoflowers inducing elastic monopoles of opposite signs, as probed with respect to $\mathbf{n}_0$. **c,d**, MSD angular-dependence of mesoflowers inducing dipolar (**c**) and quadrupolar (**d**) elastic distortions. Insets show the corresponding polarizing optical micrographs of the studied particles, taken under crossed polarizers with a 530 nm retardation plate; white double arrows indicate $\mathbf{n}_0$. Scale bars are 3 μm.

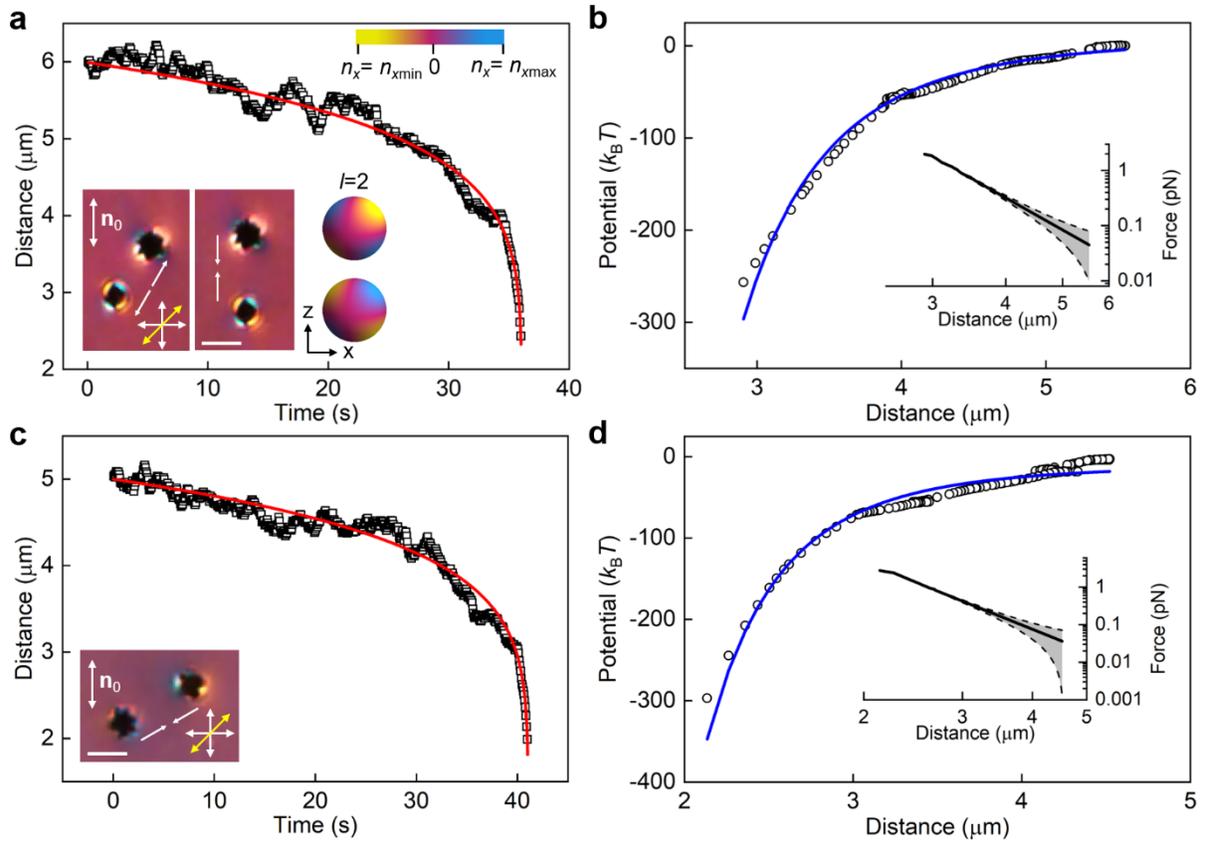

**Fig. 3 | Quadrupole-quadrupole colloidal interaction between mesoflowers. a**, Separation distance versus time for attraction between two mesoflowers inducing elastic quadrupoles with opposite signs. Micrographs in the insets show frames from movies depicting particles repelling diagonally (left) and attracting along $\mathbf{n}_0$ (right), which is expected for quadrupoles with opposite quadrupole moment signs and completely opposite from what is known for particles with the same quadrupole moment signs[17]. The direction of interaction is shown with a pair of white arrows. Schematics in the insets are visualization of the *x*-component of $\mathbf{n}(\mathbf{r})$ on the spherical surface enclosing two quadrupoles of opposite sign. Blue, yellow and magenta color indicates positive, negative, and near-zero $n_x$, as shown by the color scale bar in the top right corner. **b**, interaction potential versus distance corresponding to **a**, with the inset showing the distance dependence of force plotted using the log-log scale. **c**, Separation distance versus time of attraction between two mesoflowers inducing quadrupoles with the same signs. Polarizing micrograph in the inset shows the initial state of particles. **d**, interaction potential vs distance corresponding to **c**, with inset showing the distance dependence of force plotted using a log-log scale. The red curves in **a,c** are the best fits of the experimental data with the function $r_c(t)=(r_0^n-n\alpha t)^{1/n}$, where $n=7$ for quadrupole-quadrupole interaction; the fitting coefficients are $r_0=6.0$ μm, $\alpha=1.1\times10^3$ μm$^7$ s$^{-1}$ in **a** and $r_0=5.0$ μm, $\alpha=2.7\times10^2$ μm$^7$ s$^{-1}$ in **c**. The blue curves are the best fits of a power-law function $\propto -r_c^{-5}$ corresponding to quadrupole-quadrupole interaction potential, from which the force is calculated. Grey bands with dashes in the insets represent estimated error of the force measurement. Scale bars are 3 μm.

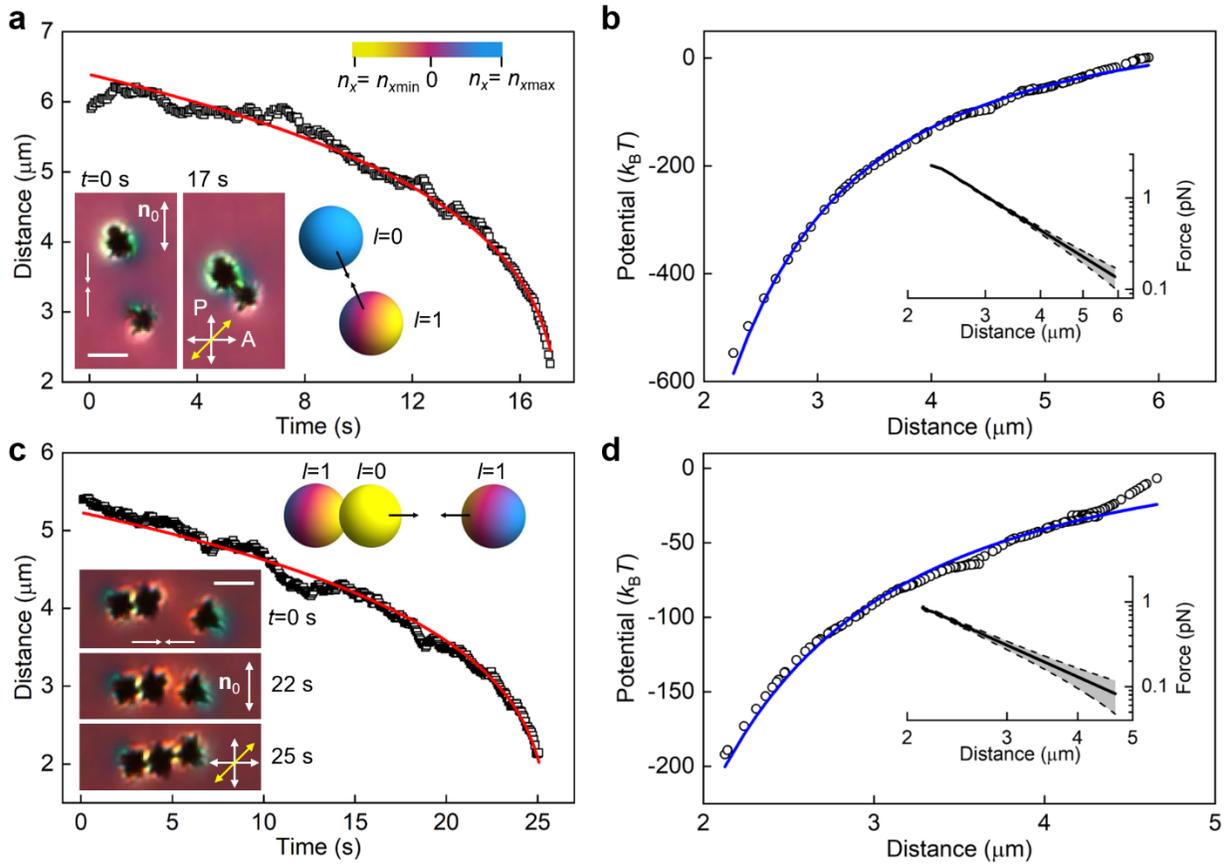

**Fig. 4 | Monopole-dipole elastic colloidal interaction between different mesoflowers. a**, Separation distance versus time of attraction between monopole-like and dipole-like mesoflowers. Insets of micrographs show particle interaction probed with polarizing optical microscopy under crossed polarizer P and analyzer A, with an additional retardation plate (slow axis marked by yellow double arrow) inserted between them. Insets of color-coded spheres are visualization of $n_x$ values at spherical surfaces enveloping the particles. Direction of interaction is shown with pairs of arrows. **b**, Interaction potential versus distance corresponding to **a**, with inset showing distance dependence of force plotted using a log-log scale. **c**, Separation distance versus time of monopole-dipole attraction between an assembly of mesoflowes and an individual one. Insets show polarizing optical microscopy movie frames and visualization of $n_x$ values at spherical surfaces enveloping the particles. **d**, Interaction potential versus distance corresponding to **c**, with inset showing the distance dependence of force plotted using a log-log scale. The red curves in **a,c** are the best fits of the experimental data with the function $r_c(t)=(r_0^n-n\alpha t)^{1/n}$ where $n=4$ for monopole-dipole interaction; the fitting coefficients are $r_0=6.4$ μm, $\alpha=24$ μm$^4$ s$^{-1}$ in **a** and $r_0=5.2$ μm, $\alpha=7.3$ μm$^4$ s$^{-1}$ in **c**. The blue curves are the best fits of a power-law function $\propto -r_c^{-2}$ corresponding to the monopole-dipole interaction potential, from which the force is calculated. Grey bands with dashes in the inset represent estimated error of the force measurement. Scale bars are 3 μm.

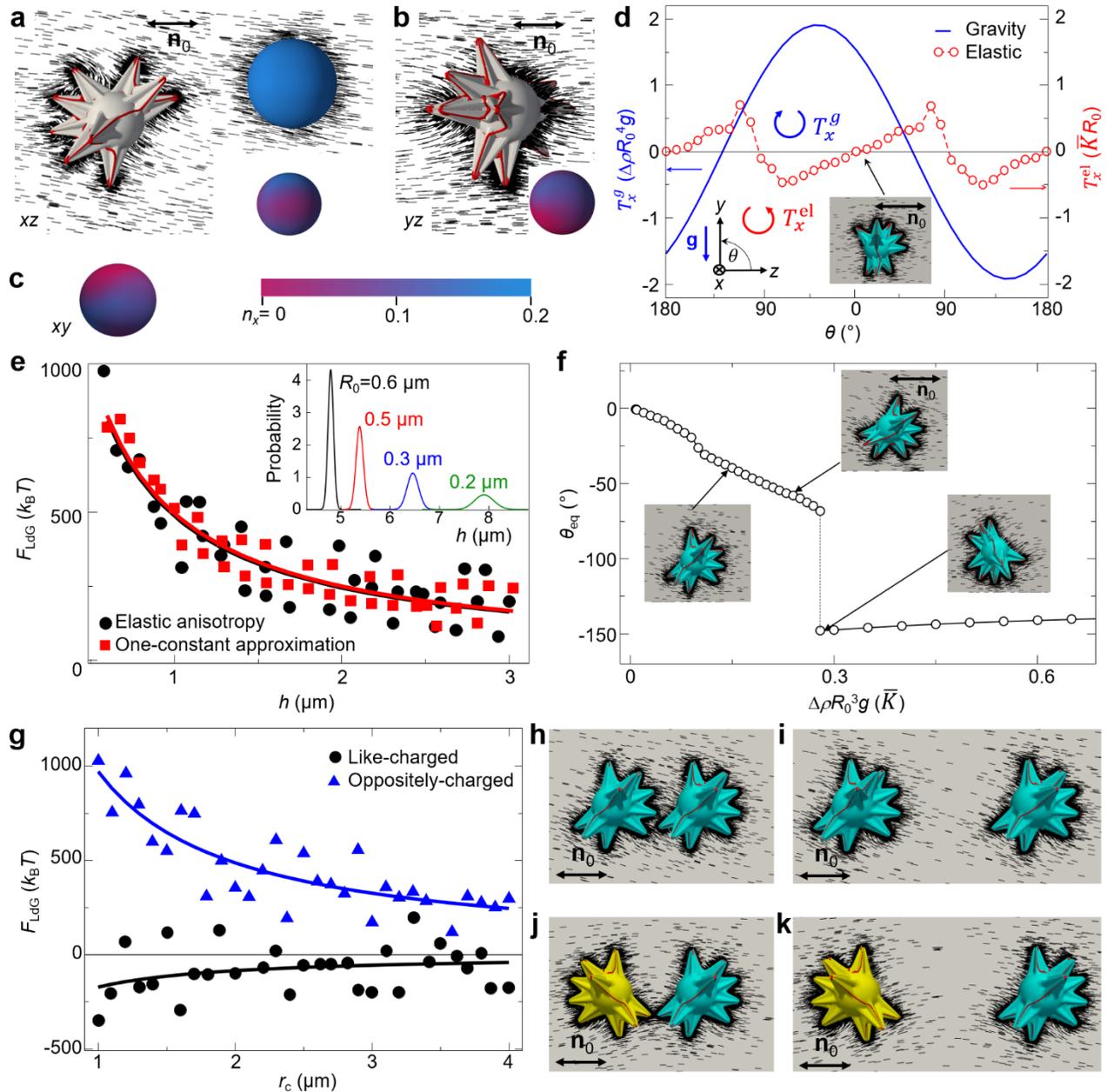

**Fig. 5 | Orientation, effective repulsion from a wall and pairwise interactions of monopoles.**
**a,b**, Director structures around a mesoflower with a dominant elastic monopole. **n(r)** is shown by black rods and defect lines are depicted as red tubes. As a comparison, **n(r)** induced by a pure idealized monopole is shown as the top-right inset of **a** whereas the bottom-right insets in **a**, and **b** show color-coded diagrams of $n_x$ on the interpolation sphere surrounding the mesoflower when viewed from different directions. The coordinate system is set so that $\mathbf{n}_0 \parallel z$. **c**, The $xy$ perspective view of the $n_x$ and the color-coded scale of $n_x$. **d**, Elastic (red circles) and gravity (blue line) torques about the center of the particle core (Supplementary Figs. 5 and 6) versus $\theta$. The gravity acts in the negative $y$-direction; **g** is the gravitational acceleration. The mesoflower is placed above a wall with fixed planar boundary conditions along $z$ and distance from the wall $h = 3.5\,R_0$.

Balance of the two torques occurs at an equilibrium orientation $\theta_{eq}$, which is plotted in **f** versus $\Delta\rho R_0^3 g$. **e**, Reduced Landau de-Gennes free energy $F_{LdG}$ versus $h$ at $\theta = -40°$. Black solid circles correspond to $K_{11} = K_{33} = 2K_{22} = 7.8$ pN and red solid squares to the one-constant approximation $K_{11} = K_{33} = K_{22} = \overline{K}$; we use $T = 298$ K and $R_0 = 0.2$ μm in the minimization. The inset in **e** shows the probability distribution $\propto \exp(-(F_{LdG} + E_g)/k_B T)$ of the particle-wall separation $h$ for several values of $R_0$ indicated next to the curves. For the Landau-de Gennes free energy we use $F_{LdG} = 5.659\, h^{-1}\overline{K} R_0^2$ (parameters extracted from the black fitting curve in **e**), and $E_g$ is the particle gravitational energy $\propto h$; average elastic constant $\overline{K} = 6.5$ pN, and $T = 298$ K. **g**, $F_{LdG}$ versus separation distance $r_c$ between a pair of like (black solid circles) and opposite (blue solid triangles) monopoles at $h = 20R_0$ $\theta_1 = \theta_2 = 40°$ for like monopoles and $\theta_1 = 40°, \theta_2 = -40°$ for opposite monopoles. Lines in **e** and **g** are fitting curves with the function $f(x) \propto 1/x$. Director structures around pairs of like-charged **h**, **i** and oppositely-charged **j**, **k** monopole particles at $r_c = 5R_0$ in **h**, **j** and $r_c = 10R_0$ in **i**, **k**.

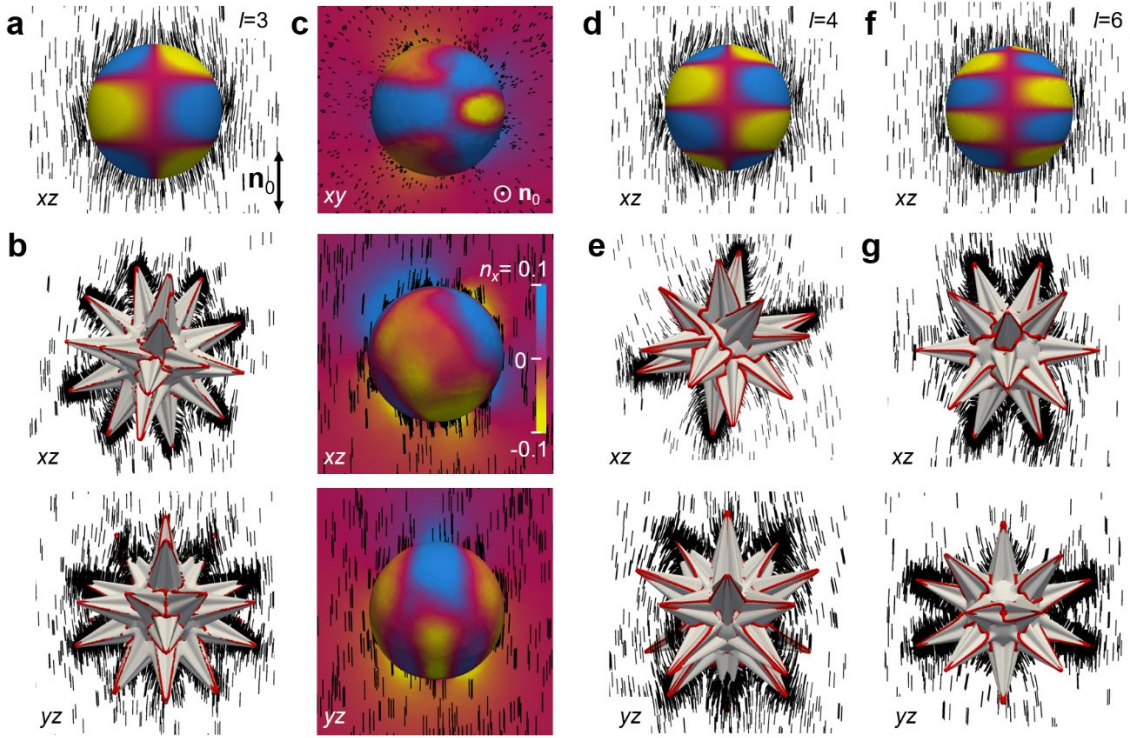

**Fig. 6 | High-order elastic mesoflower multipoles. a**, **d**, **f,** Pure elastic octupole, hexadecapole and 64-pole with corresponding **n(r)** in the cross-sectional planes, respectively, depicted on spheres using color-coded diagrams of $n_x$. **b, e, g,** Director structures around a mesoflower with dominant elastic octupole (**b**), hexadecapole (**e**), and 64-pole (**g**) contributions depicted with a perspective views on *xz* and *yz* planes. Coordinate system is defined so that **n**$_0$ ∥ *z*. **n(r)** is shown using rods and defect lines are depicted as red tubes. **c,** Color-coded diagram of $n_x$ in the *xy*, *xz*, and *yz* cross sections as well as at the interpolation spheres around the mesoflower with the dominant octupole contribution (**b**).

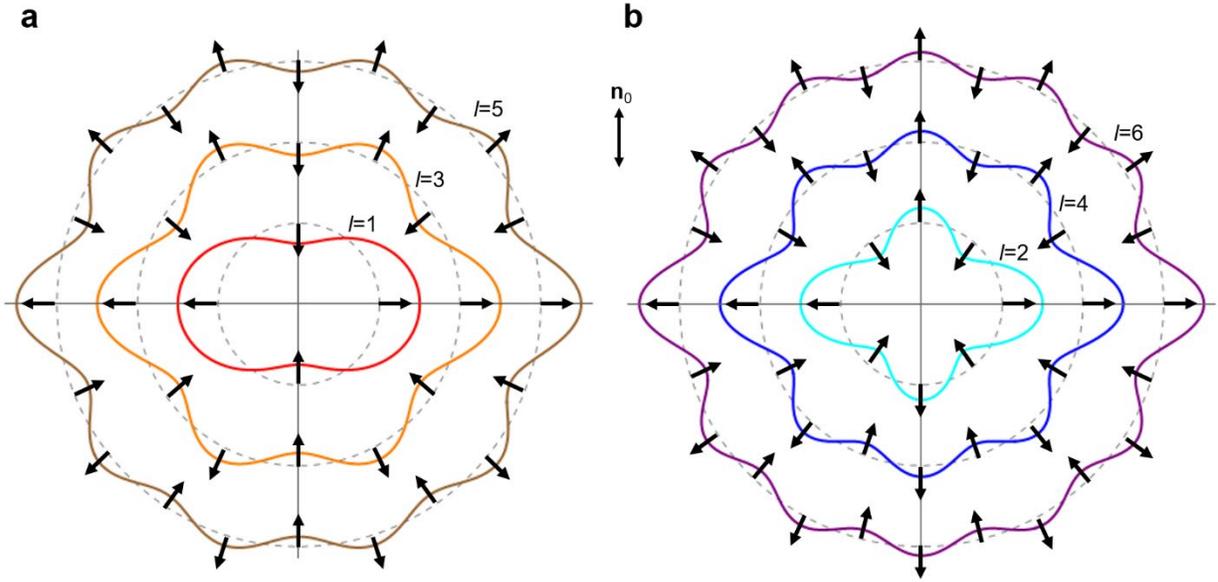

**Fig. 7 | Angular dependencies of elastic multipole pair interactions.** Colored solid lines represent the interaction potential $U_{el} \propto P_{l+m}(\cos(\theta_c))$ between two multipoles of the same order, e.g. for $l=m$. **a**, Interaction potentials between two dipoles, octupoles, and 32-poles ($l=1, 3,$ and $5$, respectively). **b**, Interaction potentials between two quadrupoles, hexadecapoles, and 64-poles ($l=2, 4,$ and $6$, respectively). The double arrow indicates the far-field director $\mathbf{n}_0$ with respect to which $\theta_c$ is measured. Dashed circles intersecting the potential plots represent equipotential lines of $U(\theta_c)=0$ for each plot. Multipole pairs mutually repel in the regions where the lines extend beyond the dashed circle, e.g. $U_{el}>0$, and attract where the lines lie within the dashed circles, e. g. $U_{el}<0$. Such radial directions of interaction are marked by the black arrows. The plots are not presented to scale.

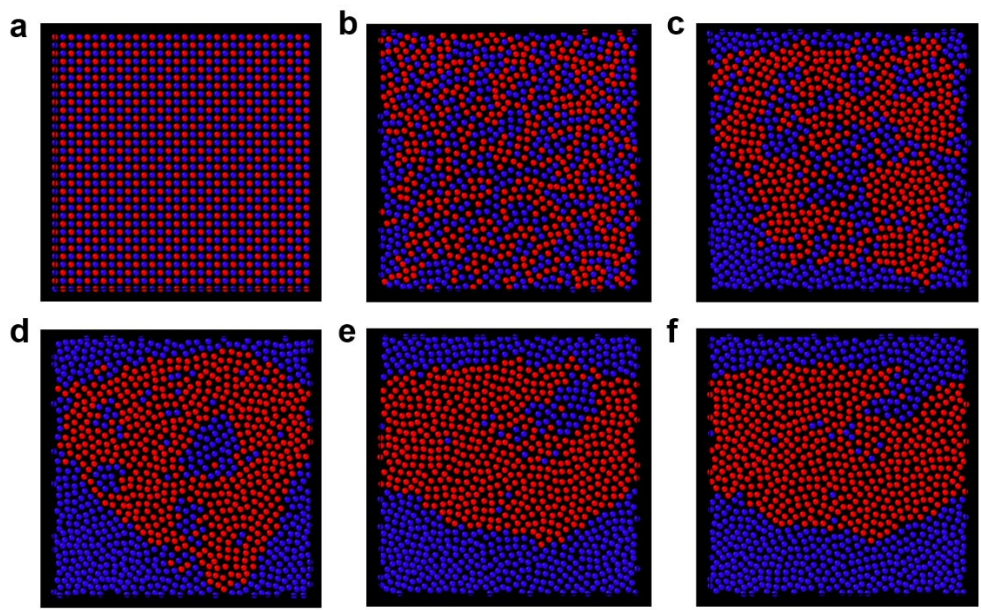

**Fig. 8 | Segregation of a binary mixture of dissimilar elastic monopoles. a**, The initial condition for molecular dynamics simulations where the colloidal particles are arranged in a NaCl-like two-dimensional square lattice. Colors of the spheres encode the monopole type: red corresponds to particles with positive elastic colloidal monopole moment, and blue to particles with negative moment. **b-f**, Snapshots sequential in time (increasing from **a** to **f**) showing spatial segregation of distinct colloidal particle.

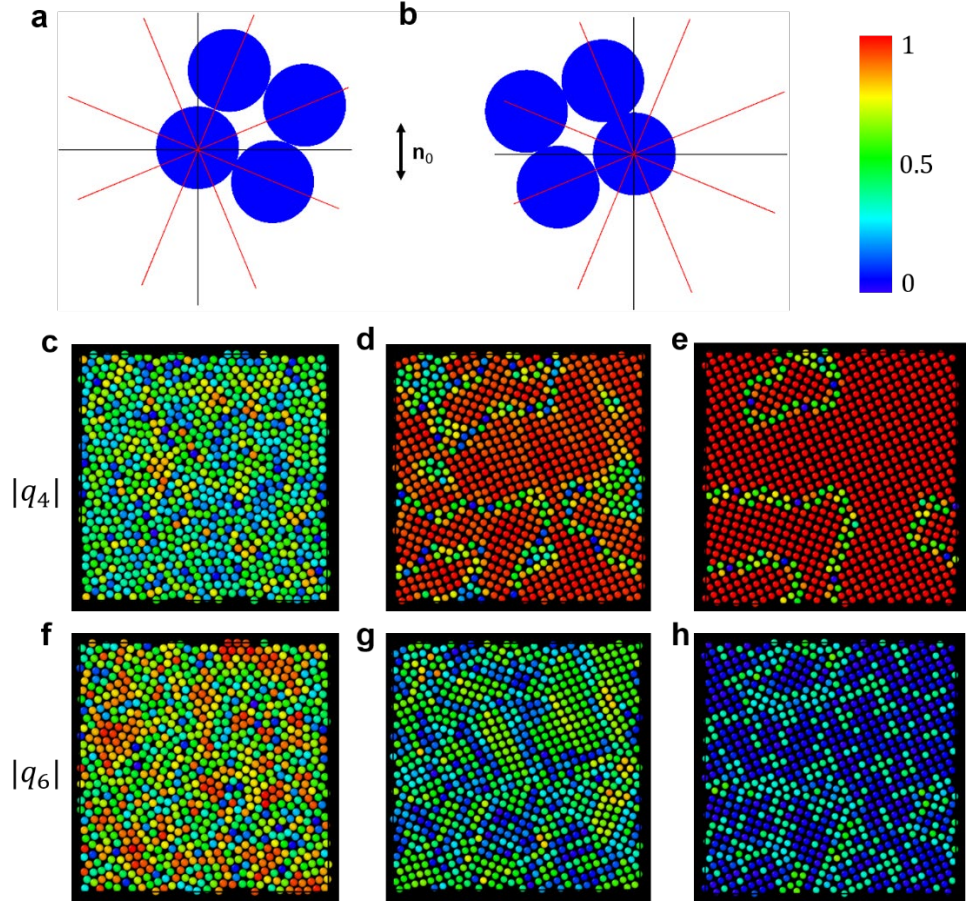

**Fig. 9 | Spontaneous crystallization of elastic hexadecapoles with grain boundaries. a,b**, Unit cells formed by the spontaneous crystallization of colloidal particles with dominant hexadecapole elastic moments. Red lines indicate the attractive directions from the underlying hexadecapole interaction potential (Fig. 7b). Black lines represent the Cartesian axes with $\mathbf{n}_0$ parallel to the vertical axis. **c-h**, Snapshots of the process of grain boundary formation. Spheres in the top (bottom) rows are colored according to the absolute value of the local quartic order parameter $q_4(j)$ (hexatic order parameter $q_6(j)$); color scale decoding the value for both of these order parameters within 0 to 1 is shown as an inset on the right side of (**b**). Panels in the same column are taken at the same time. Parameters used are $Q_4 = 7 \times 10^{-5}$, $\frac{r_{co}}{R_{eff}} = 1.5$, $\kappa R_{eff} = 2$, $\frac{A}{\bar{K}R_{eff}^2} = 1$. The system size $L_x \times L_y$ is $26 \times 26\ R_{eff}^2$ in (**c**), (**f**); $29 \times 29\ R_{eff}^2$ in (**d**), (**g**); $31 \times 31\ R_{eff}^2$ in (**e**), (**h**). The size of the spheres in all panels is shown not to scale.

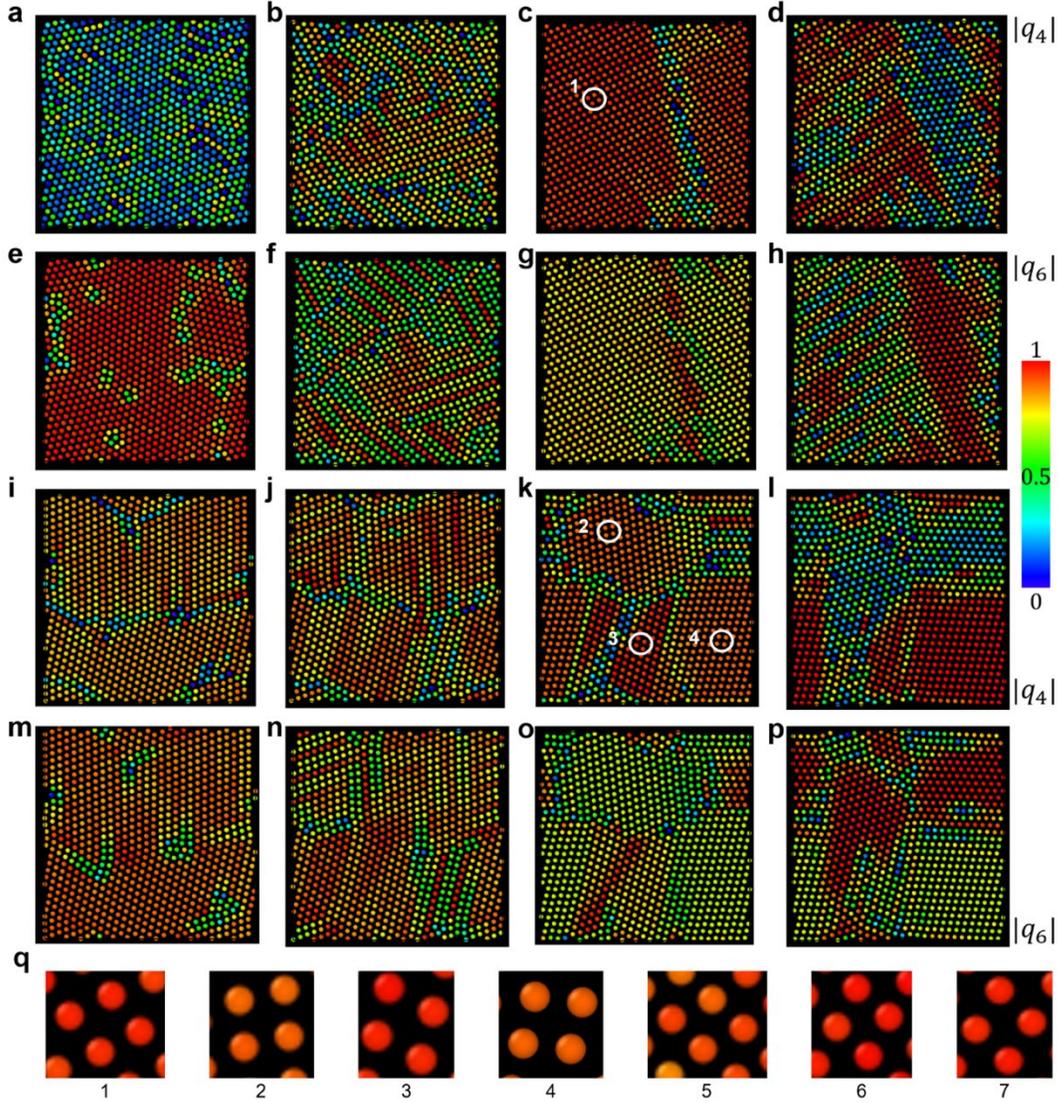

**Fig. 10 | Self-assembly of hexadecapole colloids with longer-stronger Yukawa repulsion. a-h**, Snapshots of temporal evolution of self-assembled structures. Panels in the same column are taken at the same time during the simulation; time elapses from (**a**) to (**d**). Spheres in the top (bottom) row are colored according to the absolute value of the local quartic order parameter $q_4(j)$ (hexatic order parameter $q_6(j)$) as marked on the right with the color scale as an inset. Parameters used are $Q_4 = 9 \times 10^{-5}$, $\frac{r_c}{R_{\text{eff}}} = 5$, $\kappa R_{\text{eff}} = 0.3$, $\frac{A}{\bar{K} R_{\text{eff}}^2} = 1$. The system size $L_x \times L_y = 32 \times 32\, R_{\text{eff}}^2$ in (**a**), (**e**); $L_x \times L_y = 35 \times 35\, R_{\text{eff}}^2$ in (**b**), (**f**); $L_x \times L_y = 36 \times 36\, R_{\text{eff}}^2$ in (**c**), (**g**); and $L_x \times L_y = 37 \times 37\, R_{\text{eff}}^2$ in (**d**), (**h**). **i-p**, Evolution of a system with stronger Yukawa repulsion. Panels are arranged and colored in the same way as (**a**)-(**h**). Parameters used are $Q_4 = 9 \times 10^{-5}$, $\frac{r_c}{R_{\text{eff}}} = 5$, $\kappa R_{\text{eff}} = 0.3$, $\frac{A}{\bar{K} R_{\text{eff}}^2} = 10$. The system size $L_x \times L_y = 34 \times 34\, R_{\text{eff}}^2$ in (**i**), (**m**); $L_x \times L_y = 35 \times 35\, R_{\text{eff}}^2$ in (**j**), (**n**); $L_x \times L_y = 36 \times 36\, R_{\text{eff}}^2$ in (**k**), (**o**); and $L_x \times L_y = 37 \times 37\, R_{\text{eff}}^2$ in (**l**), (**p**). **q**,

rhombic lattices formed by such colloids. The numbers correspond to those circled in (**c**) and (**k**), and Supplementary Fig. 7(**c**), and (**k**). The size of the spheres in all panels is shown not to scale.

**Supplymentary Figures:**

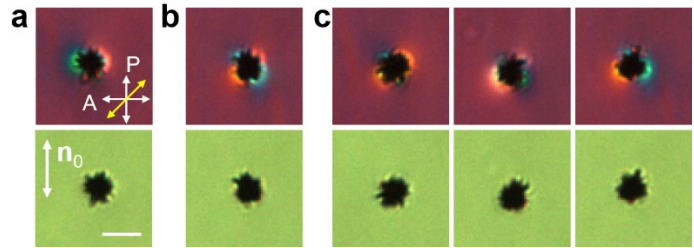

**Supplementary Fig. 1 | A single mesoflower embedded in the LC showing different metastable elastic multipoles. a-c,** Polarizing optical micrographs of the same gold mesoflower in metastable states with dipole-like (**a**), quadrupole-like (**b**) and even more complex (**c**) higher-order multipolar director distortions. The far-field alignment along $\mathbf{n}_0$ is indicated by the white double arrow; P and A show the crossed polarizations of the polarizer and analyzer, respectively; yellow double arrow shows the slow axis of a 530 nm retardation plate inserted between the polarizers. Scale bar is 3 μm.

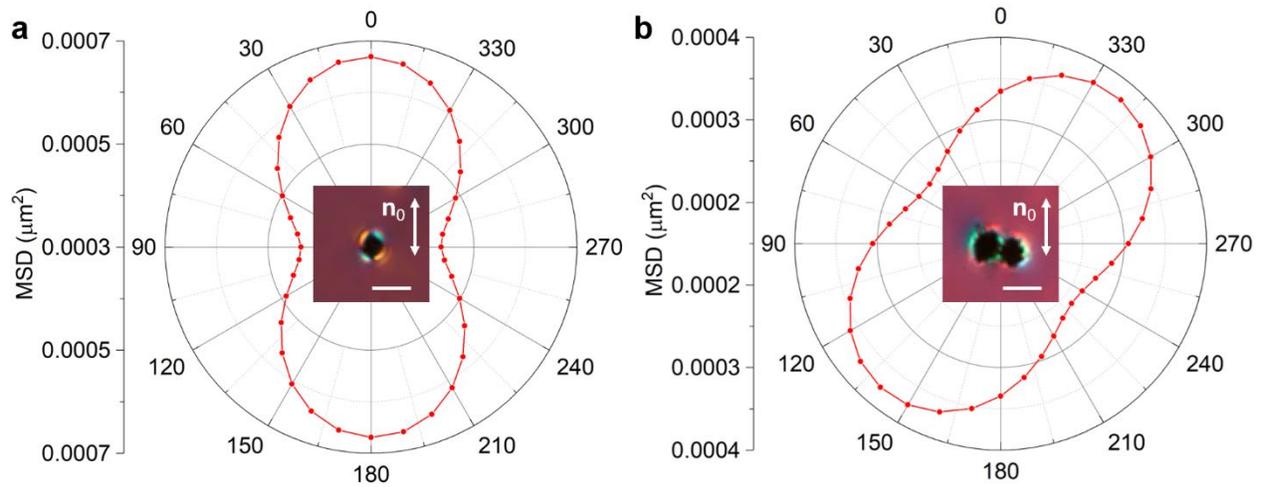

**Supplementary Fig. 2 | Anisotropic Brownian motion of colloidal particles in an aligned LC.**
**a**, Angular-dependent MSD of a gold colloidal sphere. **b**, Angular-dependent MSD of an assembly consisting of two mesoflowers. Insets show corresponding polarizing optical micrographs of the studied particles. Images are taken under crossed polarizers with a 530 nm retardation plate; white double arrows indicate **n**$_0$. Scale bars are 3 μm.

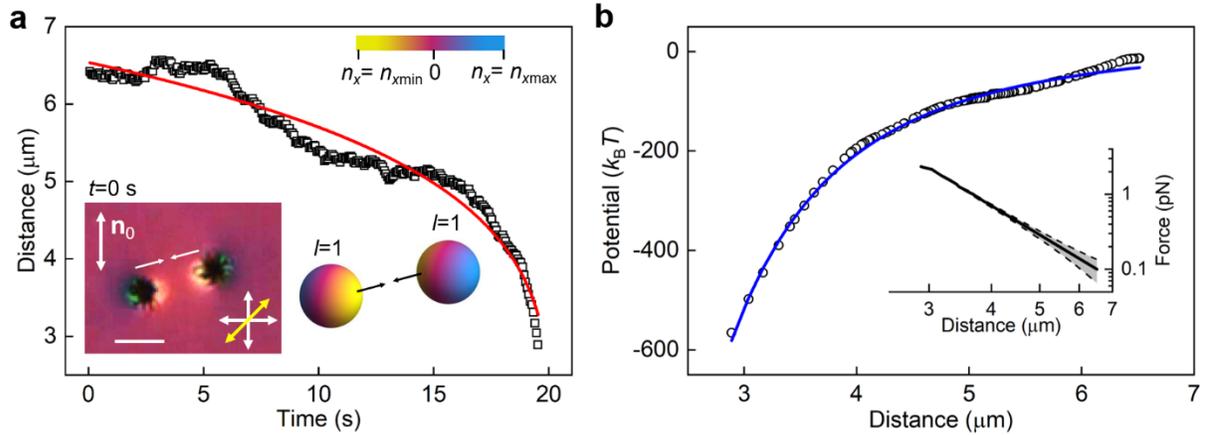

**Supplementary Fig. 3 | Attractive dipole-dipole interaction between mesoflowers. a**, Separation distance versus time for attraction between two mesoflowers inducing dipole-like structures of director field. The micrograph inset shows the initial position states of particles set by laser tweezer under polarizing optical microscopy. Schematics in the insets are visualization of the *x*-component of **n(r)** on the spherical surface enclosing two dipoles of opposite sign. The direction of interaction is shown with pairs of arrows. The red curve is the best fit of the experimental data with the function $r_c(t)=(r_0^n-n\alpha t)^{1/n}$, where $n=5$ for the dipole-dipole interaction; the fitting coefficients are $r_0=6.5$ μm and $\alpha=1.2\times10^2$ μm$^5$ s$^{-1}$. **b**, Interaction potential versus distance corresponding to **a** with inset showing distance dependence of force plotted using the log-log scale. The blue curve is the best fit with a power-law function $\propto -r_c^{-3}$ corresponding to dipole-dipole interaction potential, from which the force is calculated. Grey bands with dashes in the inset represent estimated error of the force. Scale bar in the inset is 3 μm.

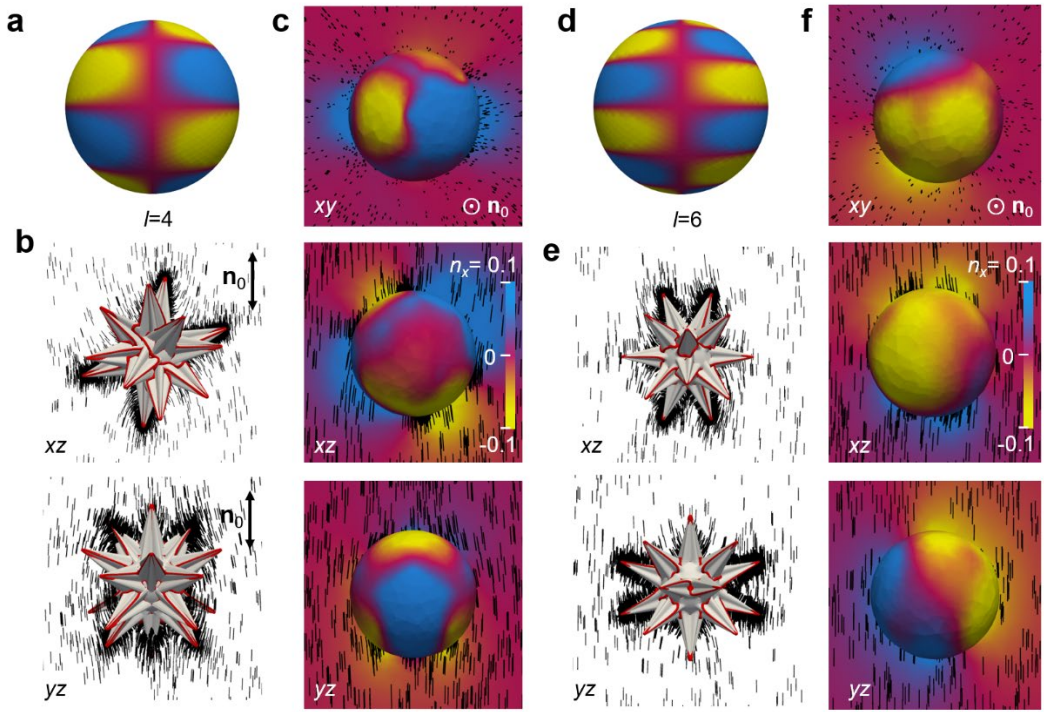

**Supplementary Fig. 4 | Hexadecapole and 64-pole. a**, **d**, Hexadecapole and 64-pole, respectively, around spherical particles with color-coded diagrams of the *x*-component $n_x$ of the director field. **b**, **e**, Director structure around a mesoflower with dominant elastic hexadepole (**b**) and 64-pole (**e**) contribution. The director fields are shown by rods and defect lines depicted as red tubes. **c, f**, Color-coded diagrams of $n_x$ in the *xy*, *xz*, and *yz* cross sections as well as at the interpolation spheres around the mesoflower with the dominant eleastice hexadecapole (**c**) and 64-pole (**f**) contribution.

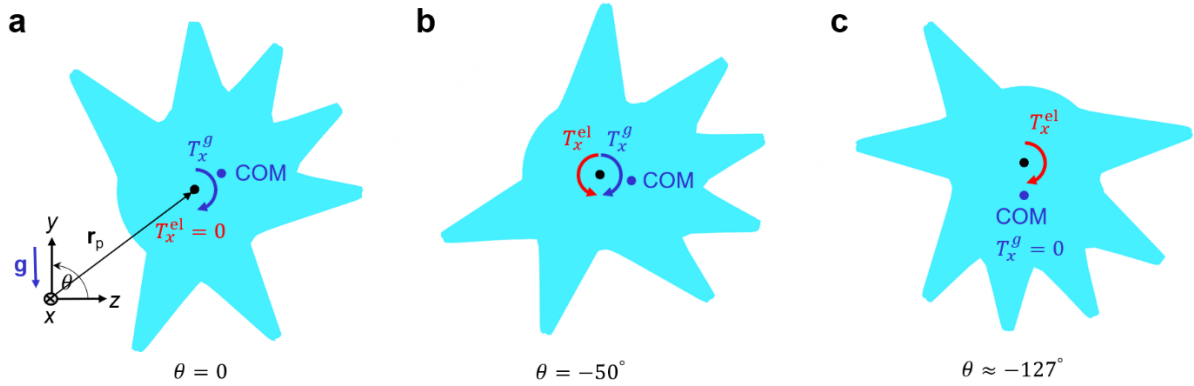

**Supplementary Fig. 5 | Cross sections of the mesoflower in Fig. 5d with a dominant elastic monopole. a**, The mesoflower at the orientation with vanishing $x$-component, $T_x^{\text{el}} = 0$, of the elastic torque about the center $\mathbf{r}_{\text{p}}$ of the spherical core of the particle, shown by the solid black dot. The $x$-component $T_x^g$ of the gravity torque rotates the particle clockwise (see the blue curve in Fig. 5c). Solid blue dot represents the center of mass (COM) of the particle; $\mathbf{g}$ depicts the direction of the gravitational acceleration. **b**, The mesoflower is tilted away from the elastic equilibrium. At this orientation $T_x^{\text{el}}$ and $T_x^g$ act in the oposite directions, conterclockwise and clockwise, respectively. **c**, The particle orientation is such that the particle center of mass and the torque pivot point $\mathbf{r}_{\text{p}}$ are on the $y$ axis, which renders $T_x^g = 0$. $T_x^{\text{el}}$ here rotates the particle clockwise (Fig. 5d).

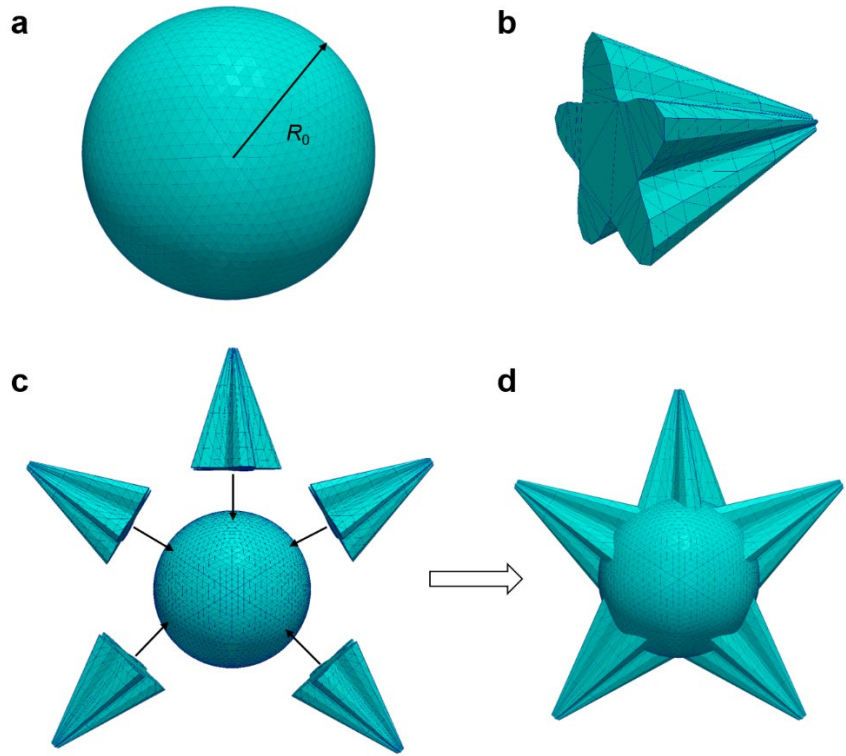

**Supplementary Fig. 6 | Schematics of the numerical generation of a mesoflower with five spikes. a**, A triangulated surface of a sphere which will be the core of the mesoflower. The sphere is characterized by its radius $R_0$. **b**, A triangulated surface of one of the spikes of the intended mesoflower. **c**, Five spikes are placed at the predefined orientations $\mathbf{\Omega}_i, i = 1,\ldots,5$, relative to the core of the planned mesoflower. **d**, Each of the spikes is translated along its $\mathbf{\Omega}_i$ towards the core and then merged with the core, by using the "*union*" function of the Gnu Triangulated Surface library, which results in a triangulated surface of the five-spike mesoflower. Other mesoflowers used in the numerical simulation are generated in a similar way with the systematically varied number and orientations of spikes.

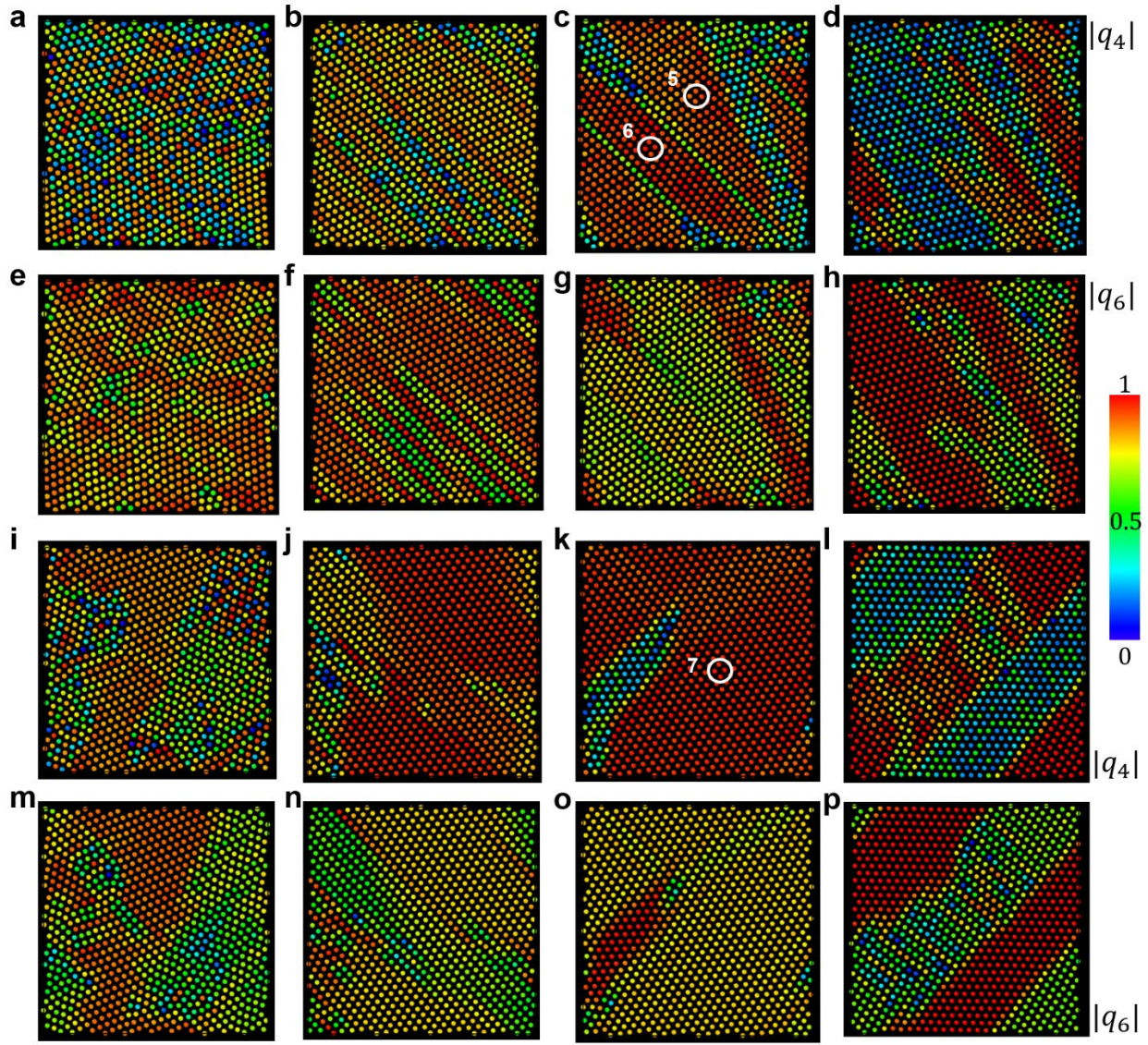

**Supplementary Fig. 7 | Self-organization of hexadecapole colloids with stronger Yukawa repulsion. a-h,** Snapshots of temporal evolution of self-assembled structures. Panels in the same column are taken at the same time during the simulation; time elapses from (**a**) to (**d**). Spheres in the top (bottom) row are colored according to the absolute value of the local quartic order parameter $q_4(j)$ (hexatic order parameter $q_6(j)$) as marked on the right with the color scale as an inset. Parameters used are $Q_4 = 3 \times 10^{-5}$, $\frac{r_c}{R_{\text{eff}}} = 5$, $\kappa R_{\text{eff}} = 0.3$, $\frac{A}{KR_{\text{eff}}^2} = 1$. The system size $L_x \times L_y = 34 \times 34\ R_{\text{eff}}^2$ in (**a**), (**e**); $L_x \times L_y = 35 \times 35\ R_{\text{eff}}^2$ in (**b**), (**f**); $L_x \times L_y = 36 \times 36\ R_{\text{eff}}^2$ in (**c**), (**g**); and $L_x \times L_y = 37 \times 37\ R_{\text{eff}}^2$ in (**d**), (**h**). **i-p,** Panels are arranged and colored in the same way as (**a**)-(**h**). Parameters used are $Q_4 = 3 \times 10^{-4}$, $\frac{r_c}{R_{\text{eff}}} = 5$, $\kappa R_{\text{eff}} = 0.3$, $\frac{A}{KR_{\text{eff}}^2} = 10$. The system size $L_x \times L_y = 34 \times 34\ R_{\text{eff}}^2$ in (**i**), (**m**); $L_x \times L_y = 35.8 \times 35.8\ R_{\text{eff}}^2$ in (**j**), (**n**); $L_x \times L_y = 36 \times 36\ R_{\text{eff}}^2$ in (**k**), (**o**); and $L_x \times L_y = 37 \times 37\ R_{\text{eff}}^2$ in (**l**), (**p**). The size of the spheres in all panels is shown not to scale.